# Modular integration of neural connectomics, dynamics and biomechanics for identification of behavioral sensorimotor pathways in *Caenorhabditis elegans*

Jimin Kim[1], Jeremy T. Florman[3], Julia A. Santos[2], Mark J. Alkema[3], Eli Shlizerman[1,2,*]
[1] Department of Electrical and Computer Engineering, University of Washington, Seattle, WA
[2] Department of Applied Mathematics, University of Washington, Seattle, WA
[3] Department of Neurobiology, UMass Chan Medical School, Worcester, MA
[*] Corresponding Author

## Abstract

Computational approaches which emulate *in-vivo* nervous system are needed to investigate mechanisms of the brain to orchestrate behavior. Such approaches must integrate a series of biophysical models encompassing the nervous system, muscles, biomechanics to allow observing the system in its entirety while supporting incorporations of different model variations. Here we develop *modWorm*: a modeling framework for the nematode *Caenorhabditis elegans* using *modular integration* approach. modWorm allows for construction of a model as an integrated series of configurable, exchangeable *modules* each describing specific biophysical processes across different modalities (e.g., nervous system, muscles, body). Utilizing modWorm, we propose a base neuro-mechanical model for *C. elegans* built upon the complete *connectome.* The model integrates a series of 7 modules: i) intra-cellular dynamics, ii) electrical and iii) chemical extra-cellular neural dynamics, iv) translation of neural activity to muscle calcium dynamics, v) muscle calcium dynamics to muscle forces, vi) muscle forces to body postures and vii) proprioceptive feedback. We validate the base model by *in-silico* injection of constant currents into sensory and inter-neurons known to be associated with locomotion behaviors and by applying external forces to the body. Applications of *in-silico* neural stimuli experimentally known to modulate locomotion show that the model can recapitulate natural behavioral responses such as forward and backward locomotion as well as mid-locomotion stimuli induced responses such as avoidance and turns. Furthermore, through *in-silico* ablation surveys, the model can infer novel neural circuits involved in sensorimotor behaviors. To further dissect mechanisms of locomotion, we utilize modWorm to introduce empirical based variations of intra and extra-cellular dynamics as well as model optimizations on associated parameters to elucidate their effects on simulated locomotion dynamics compared to experimental findings. Our results show that the proposed framework can be utilized to identify neural circuits which control, mediate and generate natural behavior.

# 1. Introduction

Neural circuits within the nervous system use rhythmic activity to facilitate coordinated body movements. Central Pattern Generator (CPG) networks are thought to generate rhythmic neural activity and motor behavior in organisms such as the locust, lamprey, *Drosophila* and the nematode *C. elegans* (1–7). However, CPG networks alone do not explain many details of sensorimotor integration and the functional pathways guiding neural activity and movement. Computational approaches which integrate the full nervous system are thus needed to identify neural circuit candidates mediating behavior. An iterative approach of discovering sensorimotor pathways via computation followed by empirical validation has the potential to discern the fundamental principles through which the nervous system and body interact (8).

In this respect, it is appealing to study the nematode organism *C. elegans,* in which inherent locomotion patterns are well observed and characterized (9). These along with environmental stimuli which lead to a change of locomotion direction provide an intriguing and well quantified model organism for locomotion investigation. *C. elegans* neuronal wiring diagram, which maps electrical and chemical neural connections between somatic neurons within its nervous system is resolved and constantly being updated across organism's sexes and developmental stages (10–20). The availability of connectome warrants searching for incorporated circuits using computational and experimental techniques. Indeed, groups of sensory-, inter- and motor-neurons have been associated with various types of locomotion including natural crawling motions (21–23), chemosensation (24–30), thermosesation (31–33) and mechanosensation (34–37). It is still not fully resolved, however, how these sensorimotor mechanisms are incorporated on a network level in *C. elegans* and what types of neural interactions lead to locomotion behaviors (38–40).

A central reason for the complexity stems from biophysical dynamics additional to the connectome. These dynamics encompass the processes representing neural responses and body bio-mechanics (22,41–46). Such processes add numerous and intricate possibilities for neural signals to flow within the routes set by the connectome. Indeed, sensorimotor integration within *C. elegans* neuron's individual responses is likely to be highly recurrent and interactive through synapses, gap junctions, neuromodulators, extra-synaptic signaling, muscles, body, proprioception, etc (46–56). While it appears as a complex system to study, the fact that these processes are coordinated during locomotion suggests that their compound effects can be inferred through the integration of individual models that emulate these processes (48,57,58). Integrating such models with varying scopes and modality warrants the development of an effective modeling framework that allows for both simultaneous simulations and testing of individual models. We therefore develop a framework allowing such a modeling approach and refer to it as *modWorm*: a modular integration modeling framework for *C. elegans* neuro-mechanics. The framework allows construction of a model as a series of configurable, exchangeable biophysical *modules* each responsible for modeling distinct dynamic processes (as depicted in Fig. 1B).

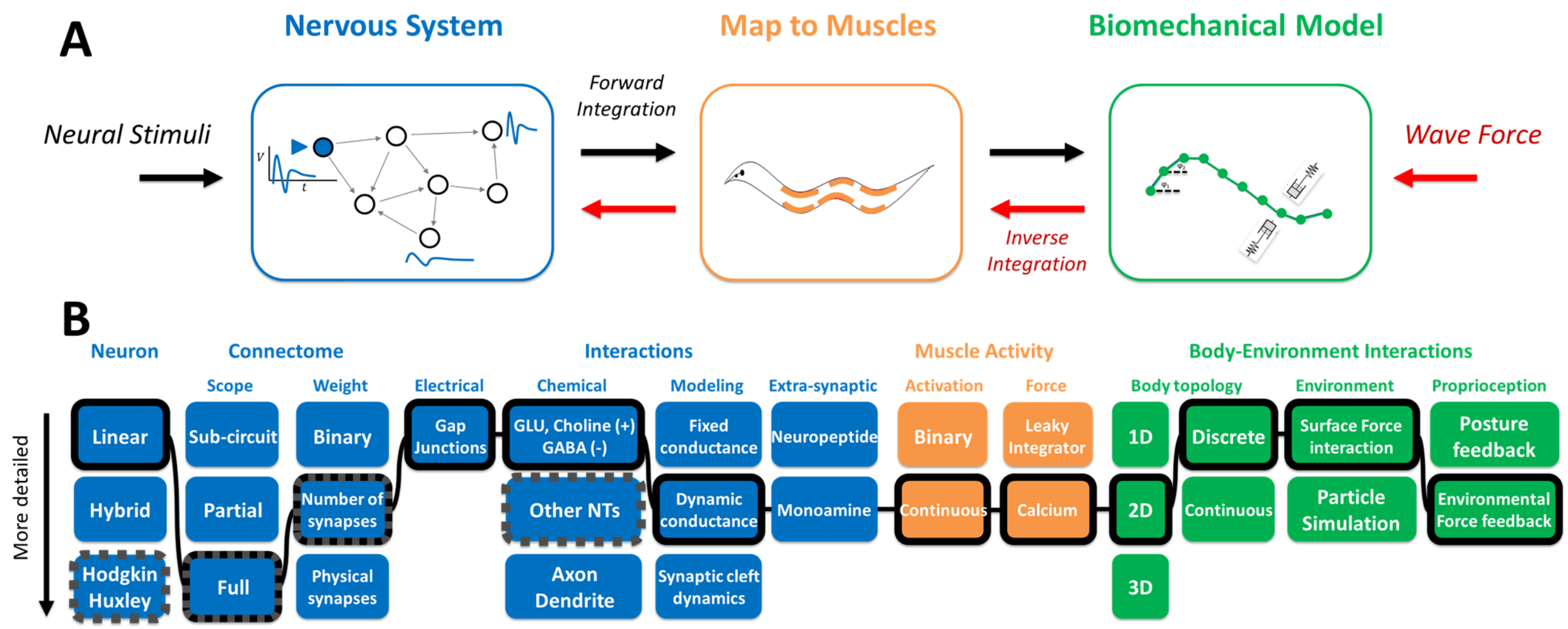

**Figure 1: Constructing *C. elegans* neuro-mechanical model**. **A:** From left to right, Modeling the nervous systems as a dynamical system encompassing the full somatic connectome including graded ion-channel and connectivity neural dynamics, mapping neural dynamics to dynamic muscle impulses and forces, mapping muscle forces to a biomechanical model that incorporates body responses and interaction with the environment. Neural stimuli are integrated forward to resolve body movements (black arrow). External forces are propagated in an inverse direction to resolve corresponding neural dynamics (red arrow). **B:** From left to right: Schematics of each model aspect as flowchart for the nervous system (blue), map to muscles (orange) and biomechanical model (green). Black highlighted boxes connected with solid lines are model aspects chosen by the proposed base model and boxes with dark gray dotted edges represent the variations considered in the paper.

Finding a combination of modules which constitute locomotory responses involves considering survey of a large pool of possible model variations. In the context of modeling *C. elegans* neuro-mechanics, this amounts to more than 50,000 variations over 13 different model aspects encompassing the nervous system, muscles and biomechanics, each with varying degrees of complexity (Fig 1B). Since it would be computationally infeasible to fully survey such large number of variations, we take a more fundamental approach where we construct an initial model comprised of *base* aspects that are generic, scalable, and unbiased by specific behaviors or experiments (i.e., no parameter fitting). These model aspects are chosen based on experimentally found anatomical and electrophysiological data as well as biophysical processes that exist in *C. elegans* (Fig 1B) (59–66). Using modWorm, the base model is constructed with a total of 7 modules incorporating simulation of the nervous system as a dynamical network, translation of neural activity to muscle forces, and mapping muscle dynamics to *C. elegans* body model.

The proposed base model of *C. elegans* integrates the full known somatic connectome with linear intracellular and non-linear extracellular neural dynamics driven by detailed synaptic transmission model similar to reduced Hodgkin-Huxley model type ion channel. The model further integrates body biomechanics and its interactions with the environment by translating the activity of the nervous system to calcium driven muscle forces which in turn generate body postures according to surrounding environment-body interactions. In addition, it implements proprioception in the form of inverse integration which modulates neural dynamics according to external forces on the body. We test and validate these model aspects by analyzing simulated neural dynamics driven by both external wave forces and neural stimuli associated with locomotion as well as studying the effects of environmental variations and proprioceptive feedback to simulated body dynamics. We also extend our studies to incorporate dynamic stimulus to investigate complex behaviors such as avoidance and turns induced by timed stimulus.

The base model permits in-place modifications to any of its modules to study its effects on neural and body dynamics. We therefore use this capability to perform *in-silico* ablations on neurons or connections and infer neural circuits involved in sensorimotor behavior associated with a stimulus. Implementation of such approaches for the study of turn response triggered by stimulus shows that we can identify neural pathways that facilitate such behavior. We use systemic ablation to also recapitulate locomotion behaviors associated with *in-vivo* ablation experiments of touch responses and perform further ablations to elucidate novel details on these experiments.

The modular structure of the base model also allows for its underlying modules to be extended or exchanged to test the refined fit of locomotion to empirical findings. We thus introduce different variations to the base model ranging from individual neuron channels, synapses to full connectome mappings to study their effects on simulated locomotion and identify variation which results in the most significant improvement in simulation quality. In addition to variations that are empirically based, we also consider model optimization through neural parameters fitting (e.g., synapse strengths) with respect to a given locomotion task and an alternative module for extra-cellular neural dynamics to elucidate their effects on simulated behavior. Testing such variations highlights the possible mechanisms of the investigated locomotion and directions in which the base model can be improved and how it can be incorporated into *in-vivo* investigations.

# 2. Related works

Models of neural dynamics and biomechanics have been introduced for several model organisms including adult and larval *Drosophila* (20,67–72), hydra (73), lamprey (74,75), leech (76,77) and rodents (78). For *C. elegans*, the availability of anatomical, electrophysiological, biomechanical, behavioral data makes the organism a suitable candidate for neuro-mechanical modeling (12,13,16,18,79–81). Indeed, several models incorporating neural and body dynamics of varying scopes have been proposed. Here we survey these models categorized into several broad approaches to highlight their contributions and differences with respect to modWorm.

**Models incorporating partial connectomes.** Several models integrating ***partial connectome data*** (e.g., sub neural circuits) with potentially biomechanics models have been introduced. Such models include methods describing ventral motor neurons as symmetric binary units which control the body of *C. elegans* segmented as discrete rods and stretch receptors (82). The model showed gaits generating forward locomotion but also locomotion instabilities when neurons dynamic properties and arrangement are slightly changed, e.g., when binary motor units are replaced by neural dynamics. Several studies also showed that pattern generators or connectome based locomotory sub-circuits (e.g., head motor neurons and ventral nerve cord) combined with body models can produce forward locomotion and basic navigation behavior such as Klinotaxis (24,83–90). While these analyses show that oscillators and sub neural circuits can produce *C. elegans* forward locomotion body postures and subsequent sensory navigation, their relation to the full nervous system and how these patterns are being generated remains unresolved. Furthermore, a unifying relation to other locomotion behaviors such as backward, turns and pirouettes remains unclear.

**Models incorporating complete connectomes.** Other lines of connectome based models introduced a dynamical model for the complete somatic nervous system (91–94). These studies showed that the full nervous system can generate dynamic rhythms even when a few mechano-sensory neurons received a constant stimulus. These rhythms, however, could not be directly associated with behaviors since additional processes of biomechanics and proprioception were not included. Inspired by the human brain project, the OpenWorm collaborative project was established in 2011 as a crowdsourcing platform aimed to develop generic bottom-up simulations of neuronal models, body and fluid simulations to lead to a full-scale *C. elegans* model (15,91,95). While there has been progress in the development of generic tools for modeling *C. elegans* and other organisms, such as Geppetto, c302 (multiscale modeling) and Sibernetic (hydrodynamic simulation) (95,96), integration of these tools into an unified framework has not yet been achieved. Moreover, incorporation of feedback between the nervous system and body such as proprioception remains unresolved.

**Inclusion of ML driven models for enhancement.** Recent works include methods inspired by Zador et al (97) adopting machine learning techniques to enhance the model and improve its simulation accuracy. These methods target parameters such as neuron polarities, connection weights/strengths and muscle-body parameters to be optimized using machine learning algorithms with respect to the established empirical data. Such methods have been applied to nervous system of both *C. elegans* and *Drosophila* to infer particular neural functions (98–101). Similar approaches have also been developed for the neuro-mechanical modeling of *C. elegans* to reproduce stereotypical behaviors such as forward locomotion (102,103). These models, however, generally include a number of compromises on model details, such as using partial connectome, less accurate first order approximation for neural dynamics, and absence of proprioception or body-environment interactions, to accommodate large scale optimization algorithms (e.g., Back-propagation through time) (104). Furthermore, their ability to simulate locomotion behaviors additional to forward locomotion such as backward, turns and pirouettes, etc are not guaranteed.

Here, we introduce a modular modeling framework: modWorm, and subsequent neuro-mechanical model of *C. elegans* that aims to address these limitations via simultaneous simulations and observations of the system in its entirety. The model integrates established biophysical processes and their approximate parameter values in *C. elegans* which encompass the complete somatic nervous system, muscles, body, and their interactions with the environment. We show that such a base model incorporates and qualitatively reproduces known forward and backward locomotion, and transitional behaviors such as turns and avoidance. By introducing experimentally driven variations and optimizations to the base model (e.g. parameter fitting), we further show the effects of each variation on simulated neural and body dynamics with respect to their base counterparts and experiments. Such an approach of implementing initial base model, followed by its variations and experimental validations can assist in making informed decisions in which the model incorporates and reflects future details and *in-vivo* findings into an expanding model of the organism (8,105–107).

# 3. Results

**3.1. Base *C. elegans* neuro-mechanical model.** We utilize modWorm to construct *the C. elegans* neuro-mechanical model. The model consists of 3 constituent models (*Nervous System, Muscles, Biomechanics*) comprised of 7 modules (Fig S1). The Nervous System Model adopts recently established dynamical model of *C. elegans* neuronal network which simulates responses of the full somatic nervous system (279 neurons) to stimuli (83,92,94). The model is based on molecular properties of neurons in *C. elegans* network and describes neural responses as combination of 3 modules: (i) graded potentials (ii) neural gap-junction connectivity (iii) neural dynamic synaptic connectivity including glutamatergic, cholinergic and GABAergic receptors. In the base model, glutamatergic and cholinergic transmitter activated ion channels are assumed to be excitatory, and GABAergic receptors are assumed to be inhibitory. These settings are configurable to changes (as we show in the section *Model variations for investigation of simulated behaviors*). The synaptic dynamics are modeled as Hodgkin-Huxley model type ion channels where the transmitter release is controlled with dynamic gating variable that is simulated alongside the membrane potential. The model incorporates local synaptic parameters determined by the connectome (i.e., number of synapses) and global biophysical conductance coefficients per type of connection. For more details on the Nervous System Model, see SM (Supplementary Materials) and (13,14).

The Nervous System Model allows for computational clamping of neurons by external current injection. It was previously observed that injection of constant current into model's sensory neurons, e.g., the posterior PLM mechanosensory neurons, evokes oscillatory neural responses in subset of motor neurons producing low dimensional attractor-like dynamics and transient dynamics with longer timescales than the intrinsic neural dynamics (93,108). Since our initial goal is to implement a base model that is generic and not specific to behavior or experiment biases, we do not fit the neural parameters of individual neuron channels and synapses. These values, however, are easily configurable through modWorm. In the section *Model variations for investigation of simulated behaviors* and Fig. 7, we perform plausible modulations to such parameters to examine the effect of additional experimental details or hypotheses (16,109–112).

We integrate the Nervous System Model with Muscles and Biomechanics models for the investigation of how the nervous system transforms its dynamics to behaviors (Fig. S1). Muscles and Biomechanics models consist of 3 modules describing (i) translation of the nervous system activity to muscle calcium dynamics, (ii) calcium dynamics to muscle forces, (iii) muscle forces to body postures according to surrounding environment-body interactions. The Muscles Model connects motor neurons excitations to muscles activations using an experimentally determined linear map (15,81) (Fig. S6). The Biomechanics model then translates muscle activations to body dynamics by resolving the interactions between the musculature of *C. elegans* (modeled as two-dimensional viscoelastic rod) and external force from surrounding fluid environment. Such an approach was proposed for investigation of eel swimming with muscles activated by external signals emulating neural excitation (113,114). We implement this model rather than the gaited model proposed for *C. elegans* or three-dimensional body model with full particle simulations (15,82,115). This is since body discretization into segments has shown to be stable under dynamic neural stimulation. Moreover, modeling the body as two-dimensional rods and body-fluid interactions in the form of damping are effective methods of simulating *C. elegans* body in a 2D plane and its surrounding environment while maintaining computational efficiency. Indeed, in the section *The effect of the environment on locomotion*, we show that such a model can emulate locomotion body postures under environmental variations similar to *in-vivo*. In previous work, statistical models, partial connectome, or synthetic muscle stimulation were used to generate body movements (82,86,115–117).

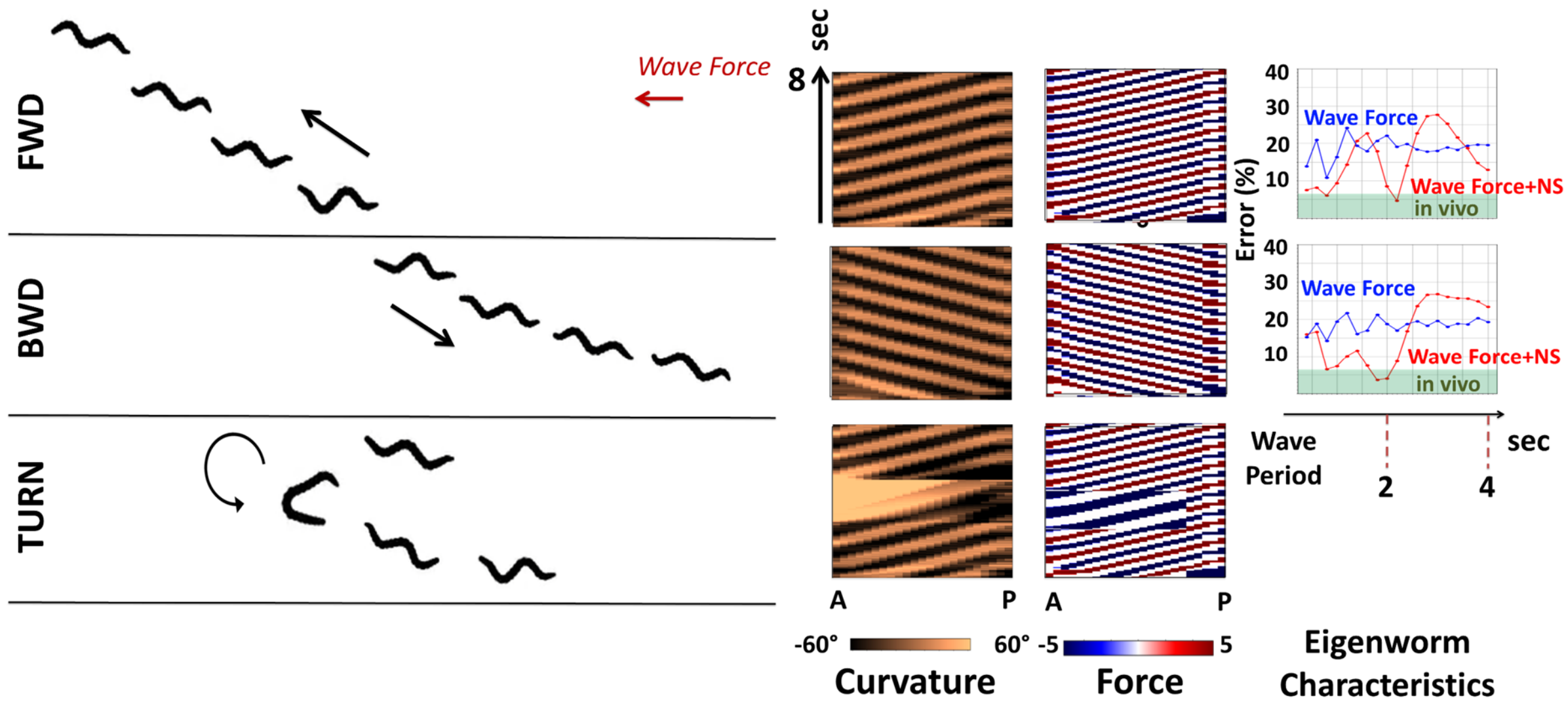

**Figure 2: Typical locomotion patterns, body curvature and force dynamics generated by three types of external wave forces, corresponding to forward, backward, 180° turn movements**. From left to right: Simulated body snapshots driven by external wave force with minimal eigenworm posture error where each snapshot is sampled every 2 seconds for forward, backward and turn dynamics (Also see SM videos), associated body curvature and muscle force dynamics ordered from anterior to posterior direction for 8 seconds simulation duration, eigenworm posture errors (w.r.t normalized posture coefficients) in the function of varying external wave force periods. Posture errors resulting from direct external wave force on the body are labeled as blue curve with 'Wave Force'. External wave force inverse integrated to resolve neural dynamics which are integrated forward to simulate body dynamics are labeled as red curve with 'Wave Force+NS'. 'Wave Force' curve results with ~20% mean error for both directions of the force and does not show preference to wave period. 'Wave Force+NS' appears to be selective to period and achieves minimal error for period ~ 2s of 4.6% (fwd) and 3.6% (bwd) within the CI of *in-vivo* worms (green band; 6.7%, P=0.01) for both directions of movement.

The implementation of 2D body model allows us to simulate the complete somatic nervous system simulation (279 neurons) for such excitation. See SM for more detailed descriptions of Muscles and Biomechanics Models.

**3.2. Locomotion and corresponding neural dynamics induced by external force.** In addition to integrating the model in a feed forward manner, we develop an inverse integration approach. Inverse integration computes the neural dynamics that would result from external forces applied to the body. The approach transforms external forces acting on the body to muscle activity and then inverts the activity to membrane potential. These potentials are then integrated forward to resolve the body posture (see SM for more details). We use this approach to validate our model for three basic locomotion patterns: forward, backward and turn movements. For each pattern we apply a sinusoidal force wave travelling along the body with variable frequency to infer neural dynamics associated with it. These neural dynamics are then forward integrated by the nervous system to generate the body posture. We then simulate the integration and compare with the locomotion characteristics of freely moving animals (see SM Videos, snapshots, curvature maps, and calcium activity, deviation in first 6 eigenworm coefficients in Fig. 2).

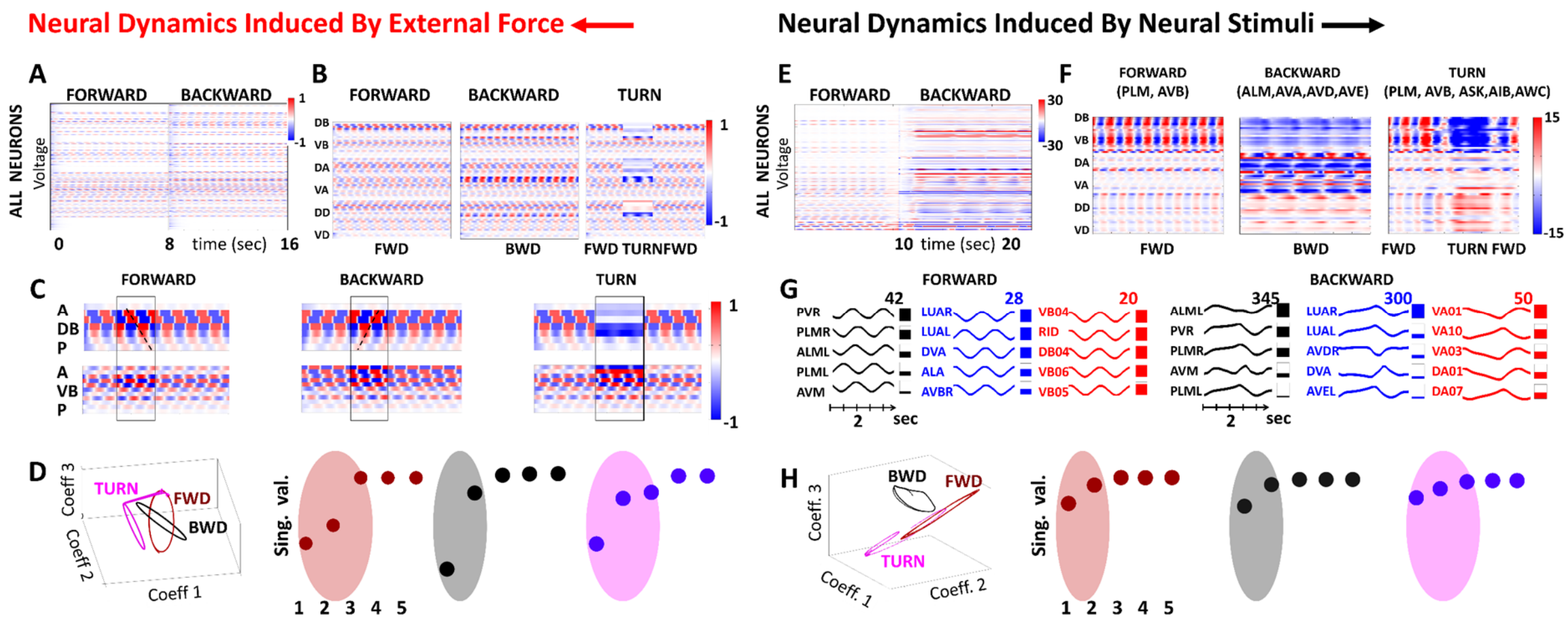

**Figure 3: Neural responses of *C. elegans* somatic nervous system to external wave forces and neural constant stimuli. A**: Color raster plot of membrane potential (difference from equilibrium) of 279 neurons inferred by inverse integration of wave forces to corresponding neural dynamics. Neural responses generated for 16 sec: 0-8 sec: spatial wave force generating forward movement; 8-9: transition; 9-16 sec: spatial wave force generating backward movement. **B**: Color raster plots of membrane potential of motor neurons for forward, backward and turn wave force profiles. **C:** Color raster plots of membrane potential of Ventral and Dorsal type B motor neurons for forward, backward and turn wave force profiles **D**: Evolution of temporal coefficients during forward, backward and turn neural responses (red, black, magenta). Temporal coefficients are associated with PC modes from SVD analysis of all three responses (i.e. projected to a common space of PC1-PC3). **E:** Color raster plot of membrane potential of optimal forward and backward current; 0-10s: forward (0.7nA into PLMR/L, 1.3nA into AVBR/L) 10-15s: transition; 15-25s: backward (2.8nA, 1nA, 0.5nA, 0.5nA into ALMR/L, AVAR/L, AVDR/L and AVER/L respectively). **F:** Color raster plots of membrane potential of motor neurons for forward, backward and turn stimulations (compare with Figure 3B). **G:** Top 5 neurons (which have largest elements in PC1 mode) from each group (sensory, inter and motor) of neurons for forward (left) and backward (right) stimuli. **H**: Evolution of temporal coefficients during forward, backward and turn neural responses (red, black, magenta); compare with Figure 3D.

We use eigenworm characteristics to compare the effect of inverse integration of external forces through the nervous system versus external forces acting directly on the body with no nervous system (9,118,119). Postures generated by the integration of the nervous system result in a close match to the coefficients of freely moving worms with preference for particular periods (4.6% and 3.6% normalized coefficient errors within 6.7% *in-vivo* error interval obtained from freely moving animals). The optimal frequency of the force that is being selected is approximately 2s. These results indicate that the response of the nervous system is shaping the external effect on the body in a nontrivial and nonlinear manner (Fig. 2).

Next, we investigate the neural dynamics associated with these locomotion patterns. In Fig. 3A, B we show membrane potential traces for the full somatic nervous system, and a group of 58 motor neurons part of the Ventral Nerve Cord (DB, VB, DA, VA, VD, DD) reported in terms of difference from the rest membrane potential. We observe that the traces are qualitatively consistent with activity patterns identified in the literature (120–123). Most of the neurons are activated during locomotion where (DB, VB) group is the most active group in forward locomotion and (DA, VA) group is the most active group in backward locomotion. When we focus on (DB, VB) neurons and order them by their physical location along the anterior to posterior axis, we find that within each period of oscillation, the activity propagates with a preferred spatial direction (Fig. 3C). During forward locomotion, membrane potential activity propagates from Anterior to Posterior (A→P) while for backward locomotion, the propagation is from Posterior to Anterior (P→A). These propagation directions are consistent with the direction of movement (124,125).

Analysis of membrane potential responses using Singular Value Decomposition (SVD) elucidates their low dimensional characteristics. The SVD method decomposes the responses into spatial neuronal population modes (PC modes) and their temporal coefficients (92,121,126). We first apply SVD to understand the representation of each individual movement to determine the number of spatial modes needed to represent each activity. The decomposition reveals that there are only a few (2-3) dominant spatial modes representing each movement similar to empirical findings (127). We thus use these modes to construct unified low-dimensional basis of spatial neuronal modes, to elucidate discriminative signatures of forward and backward movement (122). A viable candidate for such a basis is the set of the first three PC modes obtained from SVD of all motor neurons membrane potential during forward, backward and turn movements. Projection of forward and backward responses onto this basis (PC space) yields cyclic temporal trajectories which are well separated and appear to be orthogonal (Fig. 3D). When projecting the turn membrane potential dynamics onto this space we observe that the trajectory departs from the forward cycle and approaches the region of the backward cycle, highlighting the shift in neural dynamics. Notably, the PC space and the coefficients trajectories are obtained from the raw membrane potential dynamics and not from the derivative of calcium dynamics as described previously (122). This allows us to identify the PC space as a low dimensional recognition space capable of determining the type and characteristics of movements the network performs from motor neural activity.

**3.3. Neural dynamics induced by neural stimuli.** In complement to external body forces, we apply neural "clamping" to examine how these stimuli generate body movements. Our first aim is to explore movements created from simple constant stimuli where most of the neurons do not receive any input. In later investigations we extend the stimulus to be a dynamic timed stimulus (Fig. 6). To identify stimulus profile which supports coherent locomotion, we target the circuit associated with the touch response and seek for stimulations of sensory- and inter-neurons maximizing locomotion distance in either forward or backward directions (34,120,128). From neurons in this circuit, our simulations identify a subset of both sensory- and inter-neurons related to behavioral responses: posterior-touch triggered forward locomotion (sensory: PLM (0.7nA), inter: AVB (1.3nA)), anterior touch triggered backward locomotion (sensory: ALM (2.8nA), inter: AVA (1nA), AVD (0.5nA), AVE (0.5nA)), and turn movement (sensory: +ASK (0.3nA), AWC (0.6nA), inter: +AIB (0.5nA)). See Fig. S2 for illustration of optimizing the stimuli amplitudes into these neurons resulting in maximum locomotion distances. The results show that stimulation of both sensory and interneurons promote directed locomotion as reported in experimental studies and control theory analyses (129,130).

Membrane potential traces associated with neural clamping are consistent with membrane potential activity generated by external force traveling waves. We observe similar active groups of motor neurons: (DB, VB) for forward, (DA, VA) for backward, a phase change in (DB, VB) in turn (Fig. 3E), and similar preferred spatial direction in each period of oscillation. SVD analysis on membrane potential traces indicates occurrence of dimension reduction similarly to the external force case. We select the top five neurons from each neural group (sensory, inter, motor), which received the highest weight in the first PC mode, and display their membrane potential over 4 sec time (Fig. 3G). Notably, the selected neurons are those that are experimentally linked to forward and backward movements; for forward locomotion PLM, PVR sensory neurons and VB motor neurons, and for backward ALM, PLM sensory neurons and VA motor neurons. Furthermore, membrane potential response time patterns are characteristic to the two different types of locomotion examined: for forward stimulus these are clear sinusoidal membrane potential dynamics with period of ~1.8sec in- and out-of-phase oscillations and for backward stimulus these are cusp like responses over longer period of ~3.4sec with two, in- and out, phases as well. These oscillations are not present in the stimulus and are generated by the intrinsic neuronal network interactions. Projection of forward, backward locomotion and turn responses onto 3 PC modes embedding yields well separated cyclic trajectories as in the spatially traveling wave stimulation: the forward cycle is approximately orthogonal to backward cycle and turn membrane potential projected trajectory departs from the forward cycle to approach the region of the backward cycle (Fig. 3H).

**3.4. The effect of the environment on locomotion.** Experiments indicate that the environment plays an important role in shaping coordinated movement (124,131,132). We thereby explore environmental variations and their influence on the model with respect to parameters such as viscosity of the surrounding fluid and rod elasticity, which represents the ability of the rod to propagate forces along the body (133). In all variations, we fix the neural stimulus to simulate forward locomotion as in Fig. 3 and examine the effective body movement. We observe that as the environment varies, there are changes in the global characteristics of the movement as demonstrated in Fig. S10. For fluid viscosity variation, we compare the postures generated by the model to *in-vivo* experiments in which the viscosity of the surrounding agar fluid was altered by dropping a water droplet (lower viscosity than agar) (134). We mimicked such an experiment by simulating the locomotion in two different viscosity values corresponding to either agar (10mPa/s) or water (1mPa/s). We observe that for the control environment (agar), the model generates movement that features a *sine*-shaped posture qualitatively similar to that of the experiment, whereas for low viscosity (water) the same stimulation corresponds to strokes of C-shaped postures, atypical to *C. elegans* forward motion (Fig. 4A). When compared to experimental body curvatures in water, these postures have similar overall C-shaped postures as demonstrated in Fig. 4A and accompanying SM videos. These results show that a particular neural activity can support various types of behavioral responses, which are determined not only by precise choice of command neurons stimulus, but also by appropriate surrounding media for the movement.

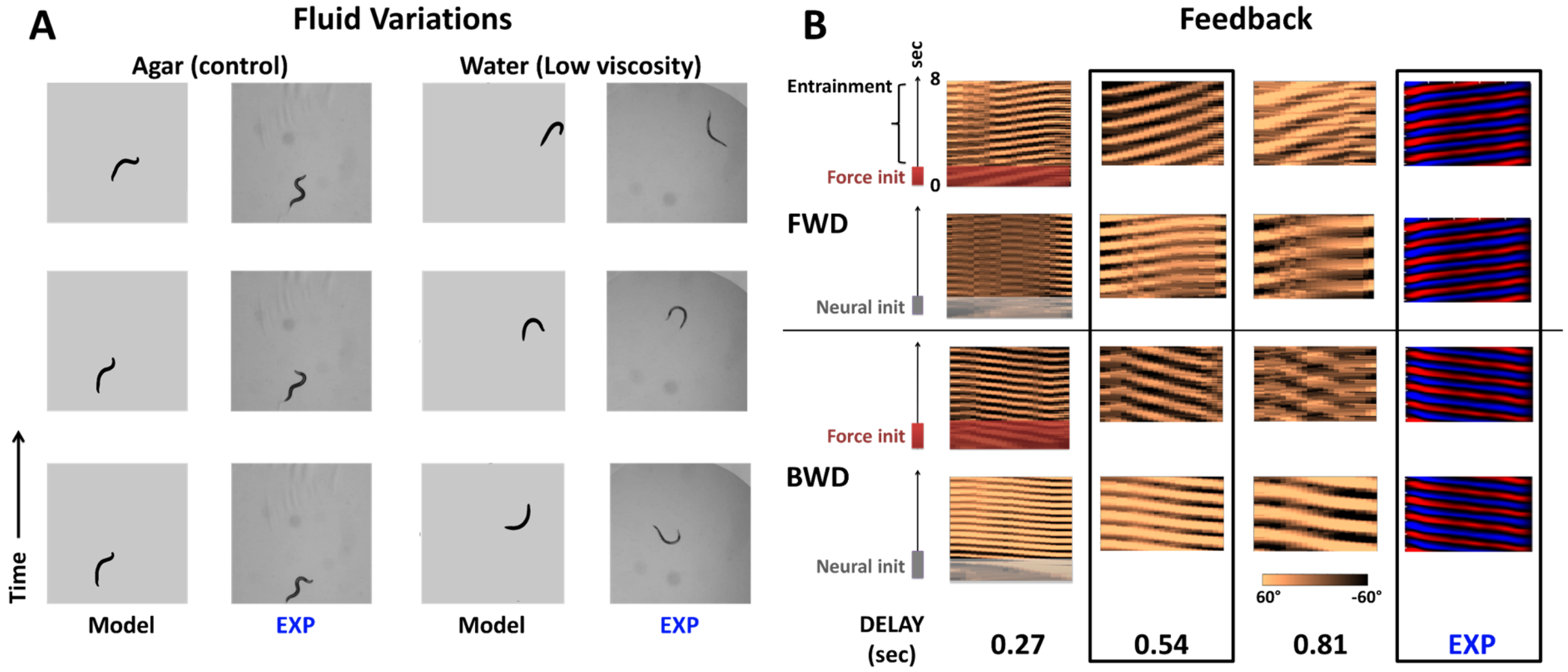

**Figure 4: Appropriate fluid parameters and proprioceptive feedback and facilitate sustained locomotion. A**: Surrounding fluid of model and experiment (134) are varied between agar (high viscosity, columns 1,2) and water (low viscosity, columns 3,4) respectively, to study their effects on locomotion. In the model, viscosity and fluid density are reduced from 10mPa/s, 1g/cm$^3$ to 1mPa/s, 0.7g/cm$^3$to emulate water droplet in experiment. For each column, 3 locomotion snapshots are taken at 1 second apart. **B**: Feedback is initiated by a wave force (rows 1,3) or neural stimuli (rows 2,4) (see also SM Videos). Columns 1-3 display various feedback time delays, modeling the environmental reaction time of producing external forces on the body, and the curvature profiles they produce and compared to column 4, experimentally recorded curvatures (adapted from (125)). Time delay of approximately 0.5 sec produces optimal forward and backward locomotion which is close to experimental locomotion (highlighted by dashed border).

**3.5. The effect of feedback (proprioception) on locomotion.** Experiments indicate that proprioception within the motor neurons circuit can facilitate locomotion and is an alternative to stimulation of command interneurons (117,131,135–139). To emulate proprioceptive feedback, we incorporate an additional inter-cellular dynamics module into the base model which outputs neural stimulation term $I^{FDB}$ based on external body forces by inverse integrating the force that acts on the body subject to a *time delay* (Fig. S1, see SM for details). Such time-delayed feedback control has been suggested to play important role shaping coherent locomotion for anguilliform swimmers (75,87,140). We then test feedback effects by initiating locomotion with external stimulation, either neural or external force stimulation. Once the feedback starts to entrain the movement, we gradually turn the stimulation off. We found that in both initiation procedures, feedback entrains the body into sustainable coherent movements in forward and backward directions, such that the body moves solely due to feedback (Fig. 4B; SM Videos). Variation in feedback delay time influences coherency and we find the delay of approximately 0.5 sec to be closest to experimentally measured patterns (Fig. 4B) (125).

**3.6. Case study of the touch responses**

**3.6.1. Validation, recapitulation and prediction of neural ablation effects.** Next, we link neural stimulations with proprioception feedback and examine locomotion responses studied in the literature for (i) *gentle* and (ii) *harsh* anterior/posterior touch. Neural and mechanical triggers for these behaviors have been identified and we use them to validate the movements generated by base model (34,37). For all four touch responses, the base model generates typical directional movement patterns (i) forward; during posterior touch neural stimulation (PLM (Gentle) and PVD+PDE (Harsh)) and (ii) backward; during anterior touch stimulation (ALM+AVM (Gentle) and FLP+ADE+BDU+SDQR (Harsh)), as shown in Fig. 5A (Control (Wild)). To test how robust the recapitulation of the response is, we perform *in-silico* ablations of random pairs of neurons (Fig. 5A; Control (Rand)) as additional controls. Indeed, random ablations do not change, on average, the characteristic velocities of the four touch responses that we considered. Next, we consider *targeted in-silico* ablations as done in prior *in-vivo* experiments and compare velocities and directions of movements with respect to the descriptions published in these experiments (34,37); See Fig. 5A. Arrows indicate *in-vivo* reported data; color of arrows indicates consistency (blue); or disparity (red).

Our ablation results are in general, qualitatively consistent with previous *in-vivo* findings (19 scenarios out of 20). For the gentle anterior touch responses (left-top in Fig. 6A) ALM, AVM we observe that ablations of (AVA) or (AVD) are the most impactful. Ablation of (AVA) nearly stops movement, while ablation of (AVD) reverses the direction to forward movement in some cases with respect to the control response of backward movement. Notably, effects of *in-silico* ablation are not simple nor additive, e.g., ablation of (AVD) alone is stronger than ablation of (AVA) + (AVD); ablation of (AVB) + (PVC) + (LUA) causes the movement to significantly slow down, while separate ablation of (AVB) or (PVC) + (LUA) permit the typical backward movement. Similarly, for the stimulation (PVD, PDE (Harsh posterior scenario); right bottom in Fig. 5A) we observe that ablation of (PVC) does not lead to a reverse in direction as in *in-vivo* experiment. For (FLP, ADE, BDU, SDQR (Harsh anterior scenario); right-bottom in Fig. 5A) ablation of (AVD) or (AVA) + (AVD) is found to reverse the direction from backward to forward. This observation is similar to the observation described in *in-vivo* experiments (34,37). In addition to the recapitulation of *in-vivo* ablation results, we perform *in-silico* ablations to further elucidate the obtained results. In the experiments, these were not performed due to technical challenges or other aspects, e.g., separate ablation of PVC and LUA was not possible in (34). Since in the model we can ablate any subset of neurons, we perform these, and further ablations and include these results in SM (Fig S 12;).

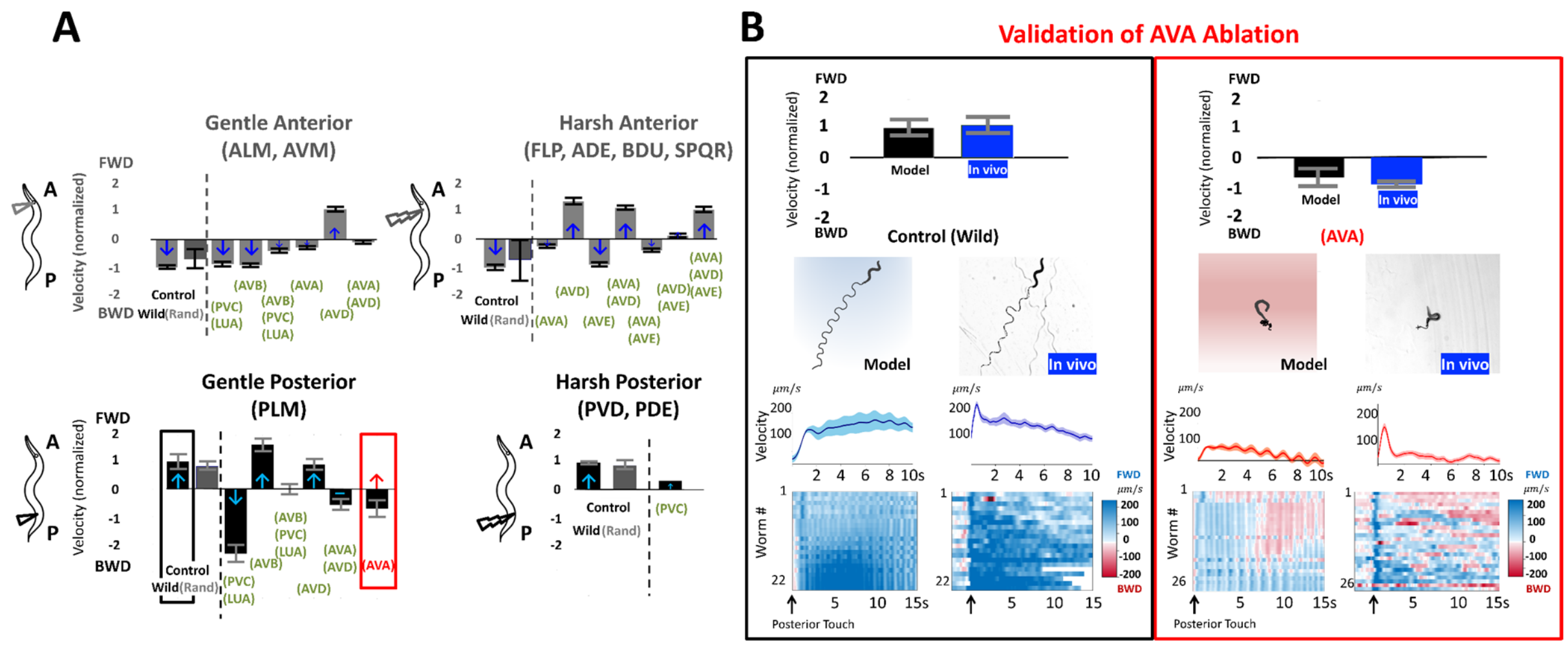

**Figure 5: Validation, recapitulation, and prediction of locomotion behaviors for touch responses. A**: Body velocities for neural stimulation associated with Top: Anterior Gentle (left) and Harsh (right) touch responses Bottom: Posterior Gentle (left) and Harsh (right) touch responses. Velocity amplitude is normalized according to the Control Wild (black label) locomotion and direction up/down are chosen as FWD and BWD direction respectively. Recapitulated velocities following ablation experiments in (34, 37) (green label) are compared to the Control Wild velocities. Arrows indicate experimental *in-vivo* observations and color indicates match of model with published experimental observations (blue: match; red: mismatch) **B**: *In-vivo* validation of Gentle Posterior Touch model prediction for ablation of AVA. Model predictions are compared with *in-vivo* assays for control (left) and AVA ablated (right). Model and *in-vivo* responses at and after stimulus onset are compared with characteristics of (top to bottom) normalized velocities, response locomotion paths, instantaneous velocities after onset, individual instantaneous velocities (each row is a worm) color plot (blue; positive velocity (fwd); red: negative velocity (bwd)).

**3.6.2 *In-vivo* validation of posterior gentle touch model prediction for AVA ablation.** For (PLM stimulation (Gentle posterior scenario); left-bottom in Fig. 5A), *in-silico* ablations provide novel predictions: while multiple ablations are consistent with *in-vivo*, the model indicates that ablation of either (AVA) + (AVD) or (AVA) significantly alter the response such that the movement is slower than control and in about half of the simulated cases invokes *backward* movement instead of the control forward movement. Since ablation of (AVD) alone does not result in significant change in the response, the model identifies AVA interneurons to have a vital role in forward movement contradicting the classical classification of AVA as a *backward* command neuron and the reported results of (34,37). Such a finding on AVA role in forward movement has been presaged by (141) and corroborated by recent experimental works (142).

Since such response was not identified in the original experiment of gentle posterior touch response, we validate the prediction with a *novel in-vivo* experiment. We use optogenetic miniSOG method to ablate AVA neurons in ZM7198 mutants and compare their responses to gentle posterior touch with control wild type N2 animals (see Fig. 5B and SM for Videos, Methods, Behavioral Assays). Gentle posterior touch was performed mechanically with hair touching the posterior part of the body. Control *in-silico* animals exhibited sustained forward movement with average instantaneous velocity of $142 \pm 22\ \mu m/s$ for the duration of at least $10s$ after the posterior touch onset. *In-vivo* control worms exhibited matching velocities, forward postures and bearing with *in-silico* control worms.

With AVA neurons ablated, *in-vivo* worms are unable to perform sustained forward movement. Behavioral assays indicate average instantaneous velocity of $19 \pm 12\mu m/s$ after the posterior touch which was in-line with the velocity of $13 \pm 22\mu m/s$ reported by *in-silico* AVA ablated. In addition, $61\%$ of tested animals (16/26) performed spontaneous backward movement for the duration of at least $1s$ (as can be seen in Fig. 5B bottom; red color corresponds to backward). The result is in qualitative agreement with recent *in-vivo* findings by (142) and *in-silico* ablation of AVA (see side by side comparison in Fig. 5B).

**3.7. Timed neural stimuli and neural ablation effects.** The touch response case study leads us to explore extensions of neural stimulus complexity and to examine the effect of timed external stimuli impulses while the worm is performing locomotion. Here we consider several case studies. The first case study is avoidance (Fig. 6A), which can be induced when forward locomotion is interrupted by ALM + AVM stimulus via anterior touch (143). In such scenarios, *C. elegans* reacts to the stimulus by stopping and reversing. We examine this scenario by initiating forward locomotion and after 6 seconds, applying ALM + AVM neural stimulation for 2 seconds. As we show in Fig. 6A, ALM + AVM neural stimulus indeed produces body dynamics similar to the avoidance behavior observed *in-vivo*, where the stimulus disturbs forward locomotion followed by initiating backward locomotion. Inspection of neural activity of motor neurons (DB neurons are A→P ordered in Fig. 6A) indicates that the stimulus induces a change in the directionality of the traveling wave of neural activity from A→P to P→A. The transition is marked through high constant activity in the anterior motor neurons.

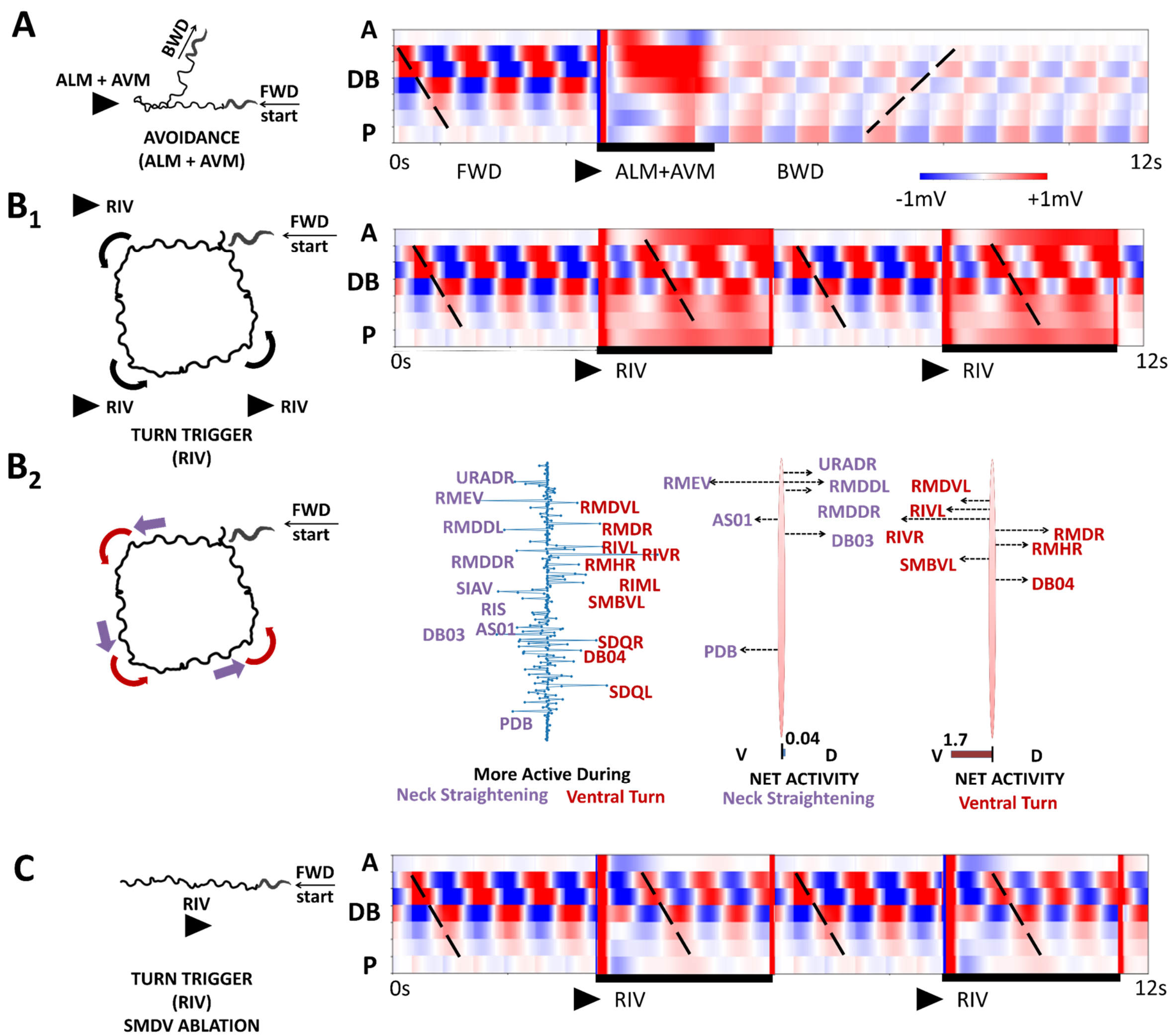

**Figure 6: Neural impulses modify basal locomotion behavior. A**: Avoidance behavior in which forward locomotion is interrupted by ALM+AVM impulse leading to backward locomotion (left: head trajectory, right: motor neurons (DB) activity). **B1**: RIV stimulus (2.7nA) is applied every 6 sec for duration of 3 sec during forward locomotion and rotates locomotion direction by 90°. **B2**: Dominant neurons and forces involved in rotation of locomotion direction (2$^{nd}$ column): (i) neck straightening-purple (ii) ventral turn-red. 3rd and 4th columns: dominant motor neurons and corresponding forces they generate in ventral and dorsal directions (3rd column - neck straightening, 4th column - ventral turn). **C**: The effect of SMDV ablation: typical RIV response (rotation by 90°) is disabled.

An additional case study that we consider is the sharp ventral turn that occurs during reorientation in the direction of locomotion (Fig. 6B). The RIV motor neurons synapse onto ventral neck muscles and have been implicated in the execution of the ventral turn (48,128). To examine the role of RIV, we stimulate

RIV neurons every 6 seconds for a duration of 3 seconds, while the worm is performing forward locomotion. As we show in Fig. 6B1, each RIV stimulus causes a sharp ventral bend of the head leading to a rotation of forward locomotion course by approximately 90° while sustaining locomotion in the forward direction. Neural activity indicates that the turn corresponds to a bias added to the membrane potential activity of oscillating motor neurons. The rotation of the body is exhibited by two posture states: (i) neck straightening followed by a (ii) ventral turn. These states are observed in experimental studies of the escape response as well (143). We investigate these states by performing SVD on neural activity in each state and identify dominant neurons associated with the activity. We then compute the force magnitude resulting from dominant motor neurons activity (Fig. 6B2). We find that during neck straightening state, dorsal and ventral forces tend to be balanced and cancel each other out, while in the ventral turn state there is a strong ventral force acting on muscle segments. Such analysis reveals neural participation on the cellular level in each state and how neural activity is superimposed to create a particular posture.

As in the touch response studies, systemic ablation can be used to further analyze the observed behaviors. Here we utilize *in-silico* combinatorial ablation of neurons to seek which neurons would be most correlated with this behavior. We select all pairs of neurons (R and L) from the group of neurons directly connected to RIV and separately ablate each pair. The analysis shows that the ablation of SMDV causes the most prominent change in dynamics, where it disables the turn behavior and causes the body to continue with forward movement, see Fig. 6C. Neural activity in the case of SMDV ablation is similar to neural activity during forward movement, suggesting that both SMDV and RIV neurons are required to facilitate a sharp turn.

### 3.8. Model variations studies

**3.8.1 Empirical variations for investigation of simulated behaviors.** Novel experimental data shows that there are additional *higher-order* properties that play role in neural activity and behavior such as spiking neurons, novel connectomics data, and extra-synaptic connections etc (16,48,109,110,112,144). The proposed base model, through modWorm, supports the incorporation of these additional properties and potential future variations. We therefore incorporate these model variations to explore the refined fits of eigenworm characteristics for the forward and backward movements in the touch response case study.

The variations we consider are (1) neurons with non-linear ion channels, (2) variation of the connectome, and (3) tyramine-gated chloride channels (LGC-55). Each of these variations was recently proposed in experiments to have a potential role in mediating locomotion (16,48,112,144). For variation (1) a group of

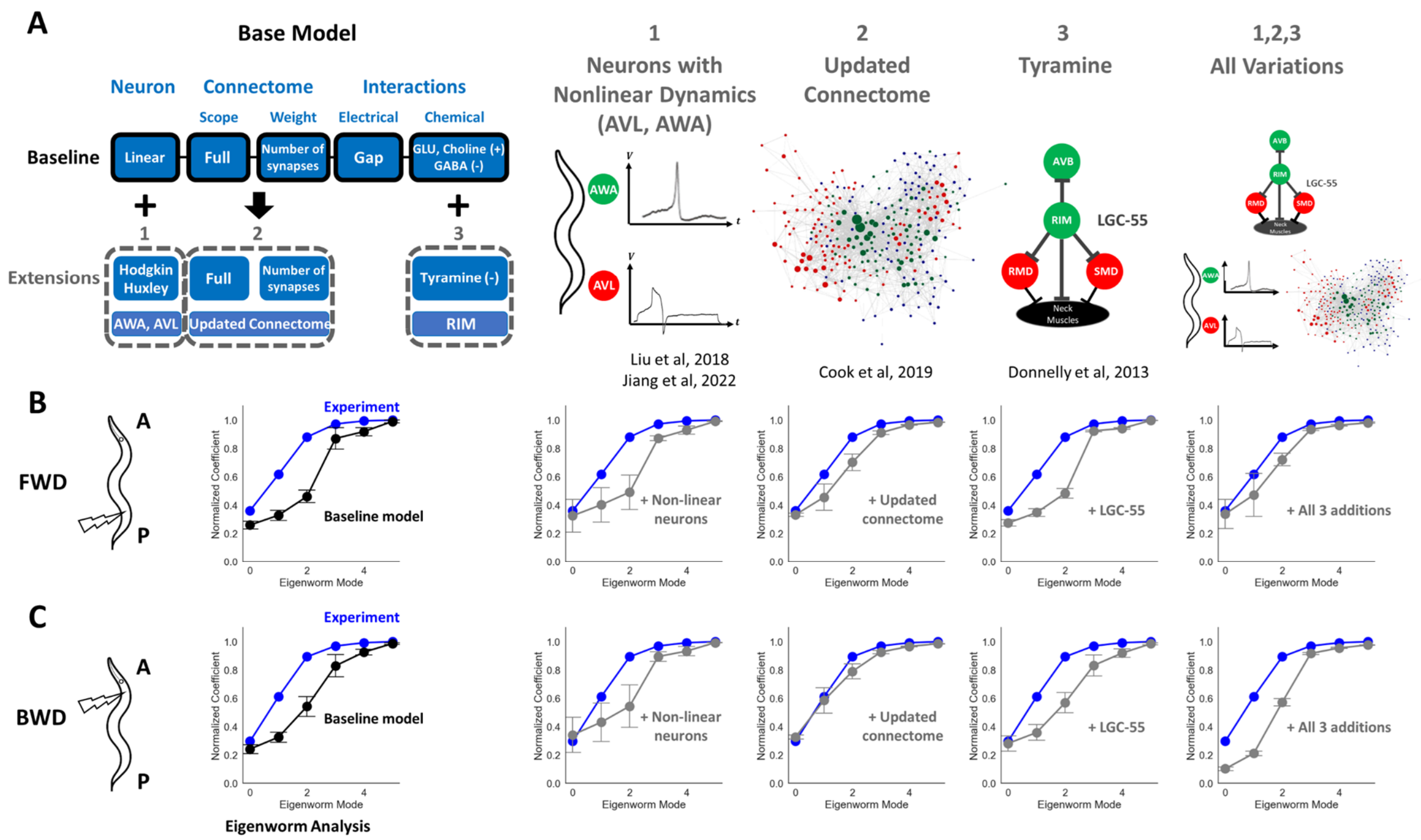

**Figure 7: Model variations and their effects on eigenworm coefficients obtained from simulated FWD and BWD locomotion. A**: Graphic illustrations of considered extensions to the model – from left to right: Neurons with "known" non-linear channels: AWA and AVL, updated connectome mappings, Tyramine gated chloride channels (LGC-55), and combination of all three additions. See Supplementary Materials for detailed implementations. **B**: Comparison of cumulative eigenworm coefficients during FWD locomotion between experiment (blue) vs base model (black) and each of the model extension (grey, N = 10, p = 0.05). FWD locomotion is simulated in the model by injecting a pulse of 3nA ($\pm 10\%$ variations each trial) of current into PLML/R followed by entrainment of proprioceptive feedback. For updated connectome, the synaptic weights are randomly varied by $\pm 10\%$ each trial. **C**: Comparison of BWD locomotion between experiment vs base model (black) and each of the model extension (grey, N = 10, p = 0.05). BWD locomotion is simulated in the model by injecting a pulse of 6.8nA and 3nA of current into ALML/R, AVM ($\pm 10\%$ variations each trial) respectively with proprioceptive feedback. Synaptic weights are randomly varied by $\pm 10\%$ each trial for updated connectome addition.

(AWA, AVL) neurons was shown to exhibit all-or-none spiking action potentials (105). The spikes are mediated by non-linear voltage-gated calcium and potassium channels and were initially observed in AWA sensory neurons, but recent experiments found such channels facilitate action potentials in enteric motor neurons (AVL) (112). While both AWA and AVL are not known to be directly associated with locomotion, we include these variations to demonstrate incorporations of additional biophysical dynamics to the base model and their possible roles in locomotion. For variation (2), we consider variation of the connectome to an up-to-date electron microscopy reconstruction that are based on the analysis of both new and published electron micrographs (16). The updated connectomics include additional connections for both gap and synaptic connectomes across the entire *C. elegans* nervous system that were previously missing or inaccurate in the old hermaphrodite connectome data (13). Lastly, for variation (3), LGC-55 is expressed by neurons that are post-synaptic to tyraminergic motor neurons. These neurons receive inhibitory signals from tyraminergic motor neurons and are known to inhibit head movements and forward locomotion during escape response (48). In modWorm, incorporating these variations into the base model can be done by modifying intra-cellular module to incorporate new individual neuron channels (variation (1)), extra-cellular modules (gap and synaptic currents) with updated connectivity weight matrices (variation (2)), or updated neuron polarity matrix (variation (3)) (see SM and Fig. S 14).

| Variation type | Movement Direction (posterior, anterior) | Mean eigenworm error | FWD eigenworm error | BWD eigenworm error |
|---|---|---|---|---|
| Static Nonlinear Synapses | (FWD, FWD) | $13.0 \pm 0.2\%$ | $11.7 \pm 0.2\%$ | $14.3 \pm 0.5\%$ |
| Optimized connectomes (w.r.t FWD locomotion) | (FWD, FWD) | $8.3 \pm 0.6\%$ | $1.9 \pm 0.2\%$ | $14.6 \pm 1.2\%$ |
| Non-linear channels (AWA, AVL) | (FWD, BWD) | $14.8 \pm 2.1\%$ | $14.7 \pm 2.7\%$ | $14.9 \pm 2.4\%$ |
| **Base model** | (FWD, BWD) | $13.1 \pm 0.8\%$ | $14.0 \pm 1.3\%$ | $12.2 \pm 1.7\%$ |
| LGC-55 (Tyramine) | (FWD, BWD) | $12.8 \pm 0.8\%$ | $13.3 \pm 1.1\%$ | $12.3 \pm 1.3\%$ |
| All variations | (FWD, BWD) | $10.9 \pm 1.2\%$ | $8.8 \pm 2.3\%$ | $13.0 \pm 0.5\%$ |
| Updated connectomes | (**FWD, BWD**) | $\mathbf{5.6 \pm 0.7}\%$ | $\mathbf{6.4 \pm 1.1}\%$ | $\mathbf{4.8 \pm 1.6}\%$ |

**Table 1:** For each model variation (row), movement directions with respect to posterior/anterior touch stimuli and normalized eigenworm coefficient error vs experiment are shown. Model variations are first sorted with respect to correct movement direction associated with posterior and anterior touch (FWD, BWD) which are then sorted with respect to mean eigenworm coefficient error in descending order.

Figure 7 and Table 1 describe the error comparison of forward and backward locomotion eigenworm coefficients (*in-vivo* vs model) obtained with base model simulations versus the models with variations. Variation (1), selected neurons having non-linear channels appear to have slightly higher error ($14.8 \pm 2.1\%$) compared to that of the base model ($13.1 \pm 0.8\%$), where the increase in error is mostly due to AVL modeled by HH-model with non-linear channels. Variation (2), updating the connectome data to the dataset published in (16) on the other hand, significantly decreased the error between the variant model and *in-vivo* coefficients ($5.6 \pm 0.7\%$). Especially the backward movement *in-vivo* coefficients were within the confidence interval ($P = 95\%$) of model obtained coefficients, indicating a closer match. From the incorporation of LGC-55, variation (3), we observe that errors for both forward and backward locomotion do not change significantly from their base counterparts ($12.8 \pm 0.8\%$). This is expected since LGC-55 was found to be primarily associated with head turning behavior during pirouette maneuver, e.g., omega turn, rather than basal locomotion (48,144). Combining all three variations decreased the error for forward ($8.8 \pm 2.3\%$) but slightly increased the error for backward locomotion ($13.0 \pm 0.5\%$). These results suggest that the effects of variations are not additive and indicate the existence of additional processes that may contribute to shaping *in-vivo* locomotion postures, and the importance of connectomes on the body dynamics.

**3.8.2. Model optimization.** Following the improvements of simulated behavior with variation (2), updated connectome data, we asked how further optimization of the connectome parameters with respect to specific behavior could affect the overall simulated behavior errors. Such task-optimized neural parameters have been successful at predicting experimental neural activity and behaviors for different organisms including Hydra, *C. elegans*, and fruit fly (73,99,115). We used a Genetic Algorithm to optimize individual synapse strengths of updated connectome data from (16) (both gap and synaptic) with respect to *in-vivo* forward locomotion eigenworm coefficients. The optimization reduced the normalized coefficient error from 5.8% to 1.9% for simulated forward locomotion (Fig. S13) (See SM for optimization procedure). To validate whether these parameters are generalizable to other types of locomotion, we apply neural stimulus associated with backward locomotion and evaluate its errors with respect to associated *in-vivo* eigenworm coefficients. This resulted in a significant increase in the error from 4.8% to 14.6% and changed the movement direction from backward to forward (Table 1, row 2). The results indicate that neural parameters (e.g., connectome mappings) optimized to specific behavior are not necessarily generalized to other behavior types or indicate biophysical parameters.

**3.8.3. Alternative module for neural dynamics.** In addition to model optimization with respect to neural parameters, we conduct variations on model dynamics equations to study its effects on simulated behaviors. We specifically modify the synaptic dynamics term $I^{syn}$ so that the transmitters release gating variable $s(t)$ is calculated with a static nonlinearity (e.g., sigmoid) dependent only on membrane potential (See SM for details). Such a variation would be more comparable to artificial neuron activation function and is a simpler approach than the original synapse model where the gating variable $s$ is governed by its own differential equation similar to Huxley-Hodgkin channel (145). The results show that while the simulated behavior under such variation has similar eigenworm coefficient errors as the base, it results in abnormality in movement properties where both posterior and anterior touch responses result in forward locomotion and the locomotion speed is significantly slower (Table 1, row 1). The results suggest that the detailed synaptic transmission model may play an important role shaping the simulated locomotion behavior closer to *in-vivo*.

## 4. Discussion

Our study provides a novel modular computational approach: *modWorm*, and subsequent neuro-mechanical model of *C. elegans* to explore the interaction between its nervous system and behavior. The proposed base model includes a total of 7 sensory-neuro-mechano-environmental modules incorporating the connectome of the full somatic nervous system, its response to stimuli and effects of neural activity on body postures and proprioception. Applications of simple stimuli in the model show that the structures of the connectome and polarities set specific movement patterns enabled by neural dynamics and biomechanics. We show that the transformations between the different constituent modules are in the form of dynamic mappings (43,45). This appears to be essential to determine whether neural activity can generate coherent body movements and cannot be resolved without incorporating the complete connectome. We introduce methods for forward integration (stimulus to muscles) and inverse integration (muscles to neural activity) allowing us to find correlation between neural stimuli and muscles and close the loop between them through proprioceptive feedback. We test the model by applying spatially travelling wave forces along the body followed by injecting constant currents into neurons in the touch response circuit. We observe that stimulation of a few neurons (sensory and inter neurons) in these circuits can generate coordinated movements consistent with direction and patterns as in *in-vivo* experiments. We then examine the effect of proprioceptive feedback and show that feedback with a time delay can entrain, smooth, and sustain locomotion initiated by neural or external force stimuli. We further test our model against previous touch response *in-vivo* experiments which used ablations to identify key neurons involved in the responses. We repeat these ablations and perform additional ablations, that were not performed in those studies or infeasible *in-vivo*, to further elucidate the roles of participating neurons in these circuits.

The model's ability to generate robust directional locomotion allows for identifying functional neural circuits and pathways associated with timed neural stimuli during locomotion. We show examples of timed neural stimuli applications during locomotion which give rise to intricate locomotion patterns and orchestrated behaviors. Specifically, we trigger behaviors such as avoidance and sharp turns through ALM+AVM and RIV neural impulses. We demonstrate that *in-silico* ablation surveys can identify neurons participating in the sensorimotor pathway of these behaviors, e.g. SMDV neuron in RIV impulse pathways.

*In-vivo* validation indicates the potential of the model to inform and complement experimental studies. In depth analysis of motion, however, shows that the characteristics of locomotion, such as eigenworm coefficients, are similar but do not precisely match with *in-vivo* coefficients (Fig. 7 Left). This is unsurprising, since the base model, by design, includes only base processes of individual neural dynamics and connections to reflect the dominant dynamic patterns and behavior. We thus utilize modWorm and introduce systemic variations to the base model to explore which additional aspects contribute to locomotion. We incorporate variations based on experimental findings such as neurons with individual non-linear channels (AWA, AVL), updates to connectome data, and synaptic channels driven by additional neurotransmitter (e.g., tyramine), that can assist in investigating which aspects contribute to better fit in locomotion metrics between *in-vivo* and the model. The results indicate that the global network properties such as connectomes may play primary roles in modulating locomotion body postures. In addition to empirical variations, we introduce theoretical variations such as task-optimized synaptic parameters and alternative model for synaptic dynamics, to elucidate their effects on simulated behaviors. Beyond the considered variations, information of the connectome, neural dynamics and processes such as monoamines signaling and neuropeptides activity modulation are rapidly become available (16,109–112,144). Moreover, modeling methodologies of biomechanics are starting to adopt three dimensions and continuous models (95,146). Investigations of the additional effects of such components could further indicate behaviors and processes that are currently not included in the base model and identify additional dynamics observed *in-vivo* such as generative spontaneous behaviors.

Beyond model optimization considered in the study, ModWorm can be used in conjunction with additional machine learning methods such as deep learning algorithms to further inform model investigations (97). For example, modWorm could be used to generate a large amount of simulation data necessary to train machine learning methods that aim to interact with the biophysical neural or body model (105,147). Leveraging the modular structure, it would also be possible for modWorm to treat parameters of biophysical modules as *learnable* (e.g., connectome weights, neuron polarity mapping) and optimize them with respect to particular neural and body task with gradient-based deep learning training methods (e.g., back-propagation through time) (148). Indeed, such approaches have already been applied to *C. elegans* and other organisms to achieve model fitting or inference of neural functions (99,102,115,149). These optimizations, however, need to be performed along with generalization requirements. Our results with connectome parameter optimization only for a subset of behaviors suggest potential simulation inaccuracies in other behaviors. Additional optimization methods (e.g., multi-objective cost function) in conjunction with gradient-based training algorithms thus can be investigated to mitigate such discrepancies. ModWorm employs community standard Python methods for constructing models (e.g., Python class) and data format (e.g., NumPy arrays). The framework is thus expected to integrate well with existing machine learning libraries (e.g., PyTorch) which adapt similar model and data structures with additional features (e.g., auto-differentiation) necessary for advanced training algorithms (150).

In conclusion, modWorm allows simultaneous simulations and variations of the nervous system, muscles and body could assist in identifying, enumerating, and classifying sensorimotor pathways. Combination of such computational studies with empirical examination and adaptation of the model may extend understanding of currently known neuromechanical functions and potentially lead to the unravelling of novel *C. elegans* brain circuits responsible for locomotion.

## Data Availability

modWorm code and generated studies reported here, manuals and tutorials demonstrating usage of the model and its variation, are available at Github repository:

**https://github.com/shlizee/modWorm**
(to be made public upon publication).

Along with the code, a blog describing the studied scenarios is available as part of the Github repository. The blog will be open, and the community will be invited to contribute to it.

**https://shlizee.github.io/modWorm/**

## Acknowledgments

The authors are thankful to the reviewers for their constructive comments. ES acknowledges the support of NSF DMS-1361145, NSF IIS-2113003 and Washington Research Fund. MA acknowledges the support of NIH RO1 GM140480. ES, JK, and JS acknowledge the support of the departments of Applied Mathematics and Electrical & Computer Engineering, the Center of Computational Neuroscience (CNC), and the eScience Center at the University of Washington in conducting this research. We thank Linh Truong and Rahul Biswas from the department of Electrical & Computer Engineering for reviewing the framework codebase.

# SUPPLEMENTARY MATERIAL

Modular integration of neural connectomics, dynamics and biomechanics for identification of behavioral sensorimotor pathways in *Caenorhabditis elegans*

Jimin Kim[1], Jeremy T. Florman[3], Julia A. Santos[2], Mark J. Alkema[3], Eli Shlizerman[1,2,*]
[1] Department of Electrical and Computer Engineering, University of Washington, Seattle, WA
[2] Department of Applied Mathematics, University of Washington, Seattle, WA
[3] Department of Neurobiology, UMass Chan Medical School, Worcester, MA
[*] Corresponding Author

## *I.* MODULAR INTEGRATION MODELING FRAMEWORK

We develop the modeling framework: *modWorm*, which we utilize to construct the base neuro-mechanical model of *C. elegans.* The framework is developed in Python and Julia programming languages and integrated with libraries such as NumPy, SciPy, DifferentialEquations.jl, and Jupyter Notebook for high performance simulations and intuitive interface (1–6). modWorm defines three model classes according to their scopes and scales: i) Module, ii) System, iii) Model. These three classes are inclusive to one another such that Module ⊂ System ⊂ Model where a Module describes individual biophysical process within a system (e.g., individual ion channel current, synaptic current), System describes a specific system within the organism (e.g., nervous system, muscles, body) and Model describes the full system in its entirety described by the integration of Systems (e.g., neuro-mechanics). Such modular modeling framework has also been adopted by popular modeling tools for nervous systems such as NeuroML (7,8).

All three model classes have two main components: parameters and dynamics equation space, which are fully customizable and describe the input-output relationship of the model. As the smallest model class, Modules serve as building blocks for larger model classes. Each Module's parameter and dynamics equation spaces are either pre-defined by the framework or created by the user. All pre-defined Modules' parameter spaces are fully configurable. A System is constructed by simply calling and combining the existing Modules to build its parameter and dynamics equation spaces (Fig. S1). Different Systems can be further combined into a Model where users can configure how inputs and outputs of Systems interact during simulations. The Model and its simulated dynamics can be further analyzed and visualized using the framework provided plotting and animation tools. The framework also allows the constructed model to be saved as a Python class object for future reusability.

## *II.* *C. ELEGANS* NEURONAL MODEL

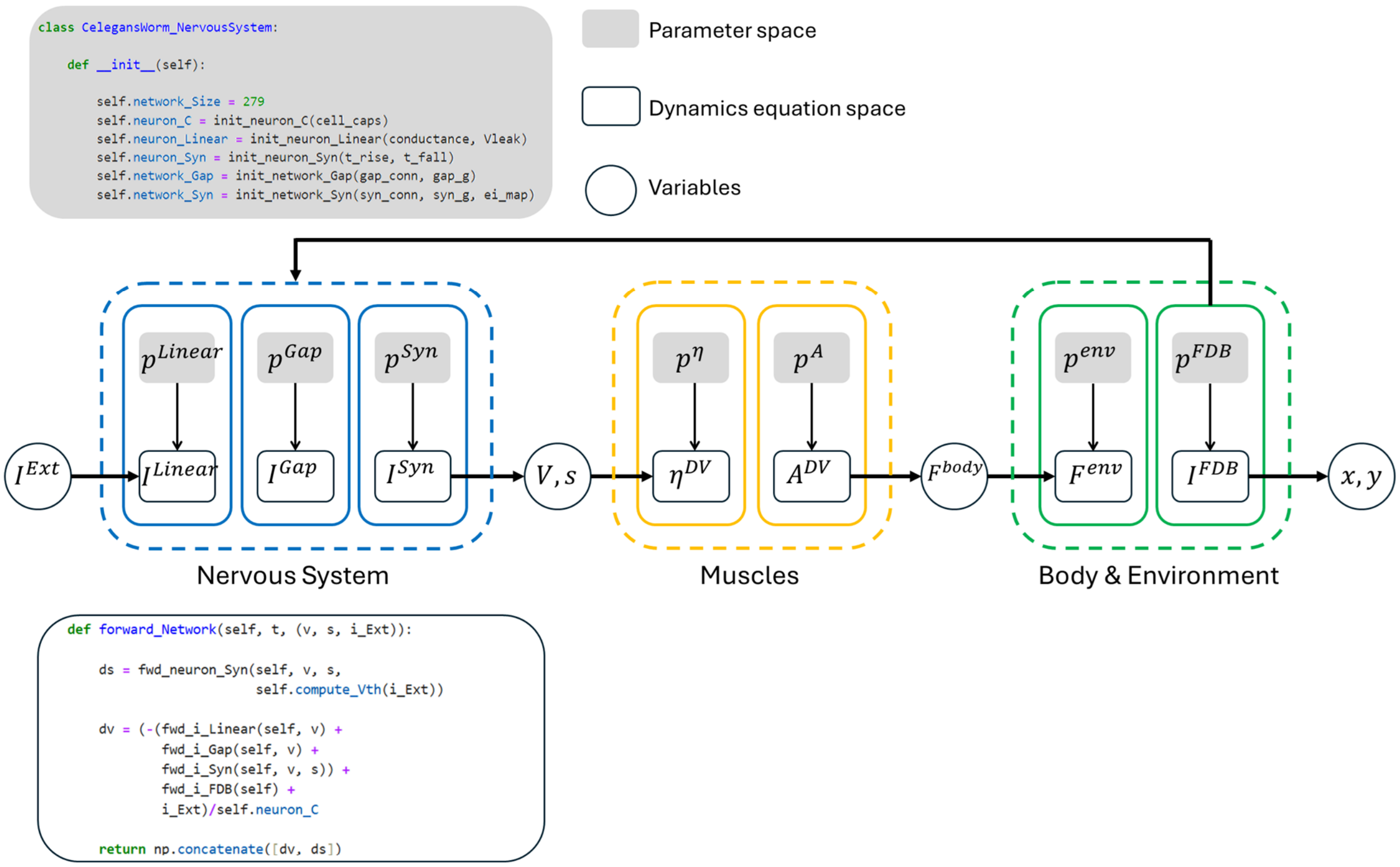

**Figure S1: Base *C. elegans* neuro-mechanical model implemented in modWorm.** The model is comprised of individual modules (solid line boxes) each with parameter and dynamics equation space. Modules are combined to form systems (dotted line boxes) which are then integrated to form a model. For the nervous system, Python code snippets outlining the example definitions of parameter and dynamics functions are shown. From left to right, neural stimuli $I^{Ext}$ is fed to the nervous system consisting of 3 modules (linear leak channel, gap, synaptic current) which outputs voltage $V$ and synaptic activity variable $s$. $V$ and $s$ are then used as inputs to muscles system to calculate muscle calcium activities $\eta^{DV}$ across dorsal-ventral muscles which are then translated to muscle activations $A^{DV}$ and outputs muscle forces $F^{body}$. $F^{body}$ is then fed to Body & Environment system to calculate corresponding environmental force $F^{env}$and proprioceptive feedback current $I^{FDB}$ and output positional coordinates $(x, y)$ for each body segment. $I^{FDB}$ is used as a feedback input to the nervous system.

To simulate neural responses, we implement the dynamical model of *C. elegans* nervous system, which consists of structural connectivity map (connectome) and equations modeling the biophysical processes of neural responses and interactions between neurons. The connectome is defined as a graph with nodes representing neurons and edges (weighted directed) representing the number of connections and their type (synaptic or gap), for more details see (9). On top of the connectome we implement dynamical equations consisting of 3 modules representing i) individual neural dynamics and two types of interactions – ii) gap and iii) synaptic (*cholinergic, glutamerigic* and GABA), as introduced in (10). In the model each neuron response is modeled by single-compartment membrane Eqn. (1) where graded potentials (leak) and gap are modeled as functions of membrane voltage $V$ and synaptic interactions between neurons are modeled with synaptic activity variable $s$ alongside with voltage $V$ as follows:

$$C\frac{dV}{dt}(t,V) = -G^C(V_i - E_{cell}) - I_i^{Gap}(\vec{\boldsymbol{V}}) - I_i^{Syn}(\vec{\boldsymbol{V}}) + I_i^{Ext} \tag{1}$$

where

$$I_i^{Gap} = \sum_j G_{ij}^g(V_i - V_j) \tag{2}$$

and

$$I_i^{Syn} = \sum_j G_{ij}^s s_j(V_i - E_j) \tag{3}$$

The parameters in the equations are total cell capacitance $C$, leakage conductance $G^C$, leakage potential $E_{cell}$, external input $I^{ext}$. Neural interactions are represented as $I^{Gap}$ and $I^{Syn}$, defined in Eqn. (2,3) and correspond to gap and synaptic junction currents respectively. The coefficients in Eqn. (2) and (3) are defined by the connectome: $G_{ij}^g$ is the total conductivity of gap junctions between neurons $i$ and $j$, and in Eq. (3), $G_{ij}^s$ is the maximum total conductivity of the synapses between neurons $i$ and $j$ modulated by the synaptic activity variable with dynamic equation:

$$\frac{ds_i}{dt} = a_r\phi(V_i;\beta,V_{th})(1-s_i) - a_d s_i \tag{4}$$

In Eqn. (4), $a_r$ and $a_d$ are the activity's rise and decay time scale coefficients respectively, and $\phi$ is the sigmoid function:

$$\phi(V_i;\beta,V_{th}) = \frac{1}{1+\exp\left(-\beta(V_i - V_{th})\right)} \tag{5}$$

Where $V_{th}$ is the network equilibrium voltage in the function of $I_i^{Ext}(t)$ that satisfies $dV(t)/\,dt = 0.\ Vth$ is thus solved at every time step to reflect time dependent $I_i^{Ext}(t)$. For more information about the derivation of the neuronal model, biophysical processes that the equations represent and units of the terms, see (10,11) and references therein.

The set of dynamical equations representing the model is 558 dimensions (2 ODEs * 279 neurons). Furthermore, these equations are stiff and thereby solved using implicit backward differentiation formulas (BDF). Specifically, we utilize VODE and CVODE solvers provided by SciPy and DifferentiEquations.jl for Python and Julia respectively when solving these equations (12,13). The system robustness also has been tested by using higher order and stochastic methods in which small external noise has been added to the parameters and verifying that the system solutions are indeed robust to these changes (10).

For efficient simulations, integration time is optimized and approximately corresponds to **actual time of the network dynamics or less**, i.e., 1 sec of integrated neural dynamics > 1 sec of actual neural dynamics. The simulation step is by default 0.01s and can be modified by the user to an arbitrary value. The integration speed can be further improved by incorporating analytic Jacobian or running the simulation in Julia as supported by modWorm. The simulations can also be run in parallel by using the ensemble simulation provided by modWorm.

## *III.* EXTERNAL STIMULATION

The term $I_i^{Ext}$(t) represents the external time-dependent current injected into each neuron. By replacing this term with different functions, we emulate stimulation of the nervous system. Fundamental stimulation is injection of constant current $I_i^{Ext} = stim_i$ into a subset of *i* neurons. Such input is generalized to step function input, where it is turned on for some time and then turned off. To avoid abrupt changes of values and hence discontinuities in the model equations, we implement smooth transition between on and off state using a sigmoid function, from one fixed value of the input to another one. The outcome is a smooth step function with typical full transition time of 2 sec. In addition, we use $I_i^{Ext}$(t) for time-dependent stimulations, such as periodic functions with various phases, periods, and amplitudes (smoothing is not required for these functions). Here, we generated the neural dynamics related to the touch circuit by using combinations of step functions as well as periodic stimuli (summarized in Table S1). Membrane voltages that are displayed and used for transformation to muscle activity (plotted in Figure 3) are voltage displacements between each neuron's membrane voltage value at time $t$, $V_i(t)$, and its resting state $V_{th}(t)$, i.e. $\bar{V}_i = V_i(t) - V_{th}(t)$.

| Stimulation Type | Neurons | Range (pA) | Behavior |
| --- | --- | --- | --- |
| Step | PLMR/L | 0-1500 | - |
| Step | PLMR/L, AVBR/L | 0-1500, 0-2500 | FWD |
| Step | ALMR/L | 0-5000 | - |
| Step | ALMR/L, AVAR/L | 0-5000, 0-2000 | - |
| Step | ALMR/L, AVDR/L | 0-5000, 0-2000 | - |
| Step | ALMR/L, AVER/L | 0-5000, 0-2000 | - |
| Step | ALMR/L, AVAR/L,AVER/L, AVDR/L | 0-5000, 0-1000, 0-1000, 0-1000 | BWD |
| Step | PLMR/L, AVBR/L, ASKR/L, AWCR/L, AIBR/L | 0-1500, 0-2000, 0-2000, 0-1000, 0-1000 | TURN |
| Periodic $1 + a\sin(p2\pi/t)$ | PLMR/L | a: 0-3000<br>p: 0.5-4 | FWD, BWD |
| Periodic $1 + a\sin(p2\pi/t)$ | ALMR/L | a: (0-3000) * 2.3<br>p: 0.5-4 | FWD, BWD |
| Periodic $1 + a\sin(p2\pi/t)$ | PLMR/L, ALMR/L | a: (0-3000)<br>p: 0.5-4 | FWD, BWD |

**Table S1:** Summary of step and periodic stimuli (neurons, ranges of amplitudes and parameters) applied to neurons for investigation of the touch response.

## *IV.* STIMULI OPTIMIZATION

To determine whether step stimulations can induce coordinated movements we employ both gradient based approach (line search) and direct optimization (survey of a range of parameters) for stimulus amplitude. The gradient method is used to identify which neurons and range of stimulations can produce locomotion in a particular direction. Direct optimization is then used to accurately map these ranges. For estimation of the locomotion coordination and direction we use a simple measure of average distance passed by a segment on the body model (tail or head). We find the optimization function to be non-convex with multiple minima, which makes it difficult to find global minimum and requires multiple trials by optimization procedure to find sufficient minima using the line search.

We set the algorithm to search for step currents that optimize the distance of locomotion into a particular direction and the total squared current amplitude $\|\vec{a}\| = \sqrt{\sum_i a_i^2}$. We focus on touch circuit and examine stimuli into neurons (sensory, inter neurons) that were found experimentally to be involved in this circuit (14,15). For forward locomotion, line search shows that the space could be spanned by two currents: PLML/R and AVBR/L (with identical current into both left and right neurons). We compute the space of currents in the range of (0, 1.5nA) for PLMR/L and in the range of (0, 2.5nA) for AVBR/L, where 1nA = 1000pA. We find the minimum with lowest amplitude at PLMR/L = 0.7nA and AVBR/L = 1.3nA (total squared amplitude is $0.73nA^2$), see Fig. S2 Left. For backward locomotion, line search identifies four currents that play a critical role: ALMR/L, AVAR/L, AVDR/L, AVER/L. We fix 2 two-dimensional spaces and first find optimal parameters in AVDR/L, AVER/L space (Figure S2 Right). This resulted in AVDR/L = 0.5nA, AVER/L = 0.5nA according to our minimum with lowest amplitude. We then follow with a search for a minimum in ALMR/L, AVAR/L space. We find the minimum with lowest amplitude to be at ALMR/L = 2.9nA and AVAR/L = 1nA (total squared amplitude is $0.78nA^2$), see Figure S2 Right.

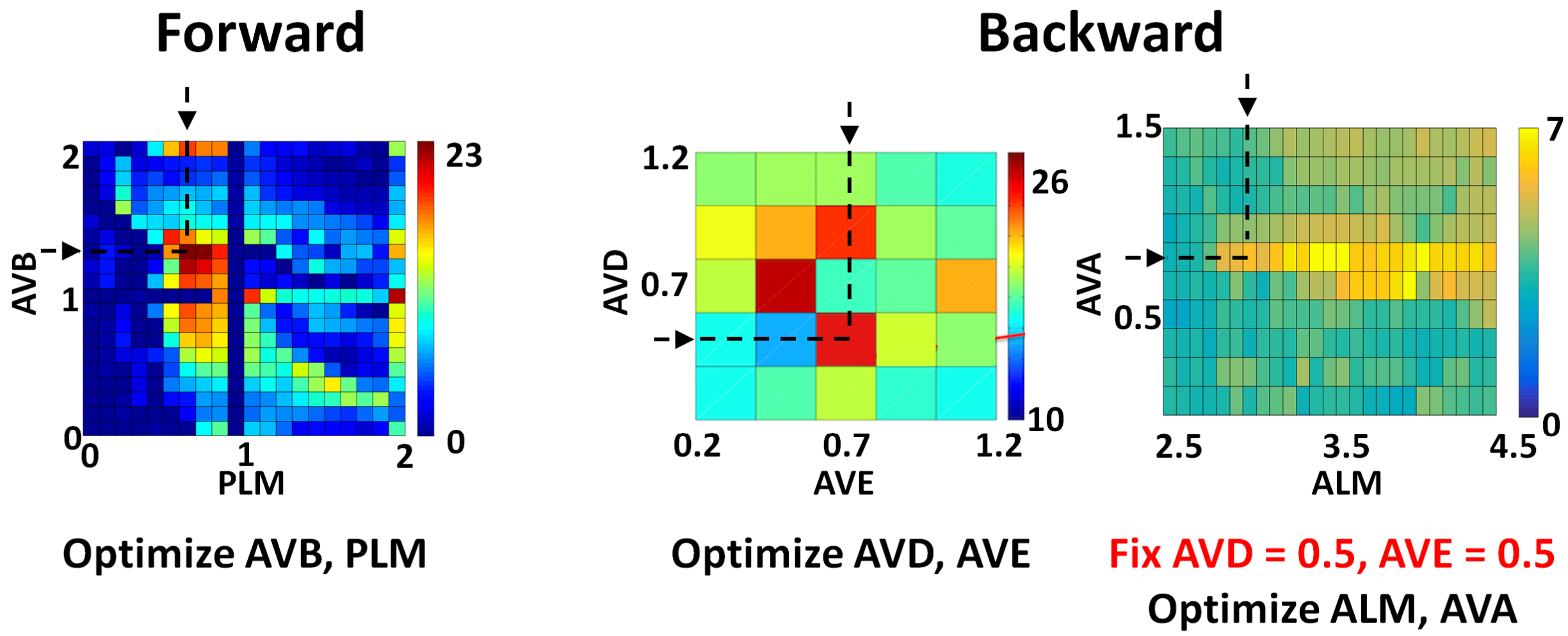

**Figure S2:** Optimization for forward and backward locomotion. Left: Optimization over two-dimensional parameter space (AVB and PLM) which results in the maximum forward distance traveled. Right: Optimization over four-dimensional parameter space (ALM, AVA, AVD, AVE) which results in the maximum backward distance. The parameter space is split into two spaces (AVD, AVE) first and then (ALM, AVA). Color corresponds to distance passed during locomotion. 1000pA = 1 is used for stimulus amplitude unit.

For turn dynamics we add the currents of AIBR/L, ASKR/L, AWCR/L neurons to forward optimal stimulus and employ a direct search to find the lowest currents that will modify typical forward neural and behavioral responses. Multiple current settings change responses in various ways. We find the lowest and most significant change when we set ASKR/L = 0.3nA, AWCR/L = 0.6nA, AIBR/L = 0.5nA and use it in Fig. 3.

## *V.* DIMENSION REDUCTION OF NEURONAL DYNAMICS

We investigate the patterns of the *C. elegans* neural network activity during forward, backward locomotion and turn by performing singular value decomposition (SVD) on neuronal activity data (voltage difference from equilibrium) (see Fig. 3 and Table S1). We focus on analyzing ten-second simulations which were sufficient to describe the representative dynamics associated with the stimuli. In all cases, the first second of simulation data was removed to ignore the effects of the transitional forced perturbation. The SVD is applied on simulation data, which is a matrix $D$ of dimensions $n \times t$ (each row corresponds to $i$th neuron, and column corresponds to a time sample $t$) and its elements are $\bar{V}_{i,t}$ (16,17). The raster plots for matrices $D$ for different types of neuron groups (complete connectome, dorsal motor neurons, ventral motor neurons) and different external inputs (FWD, BWD, TURN) are shown as color plots in Fig. 3.

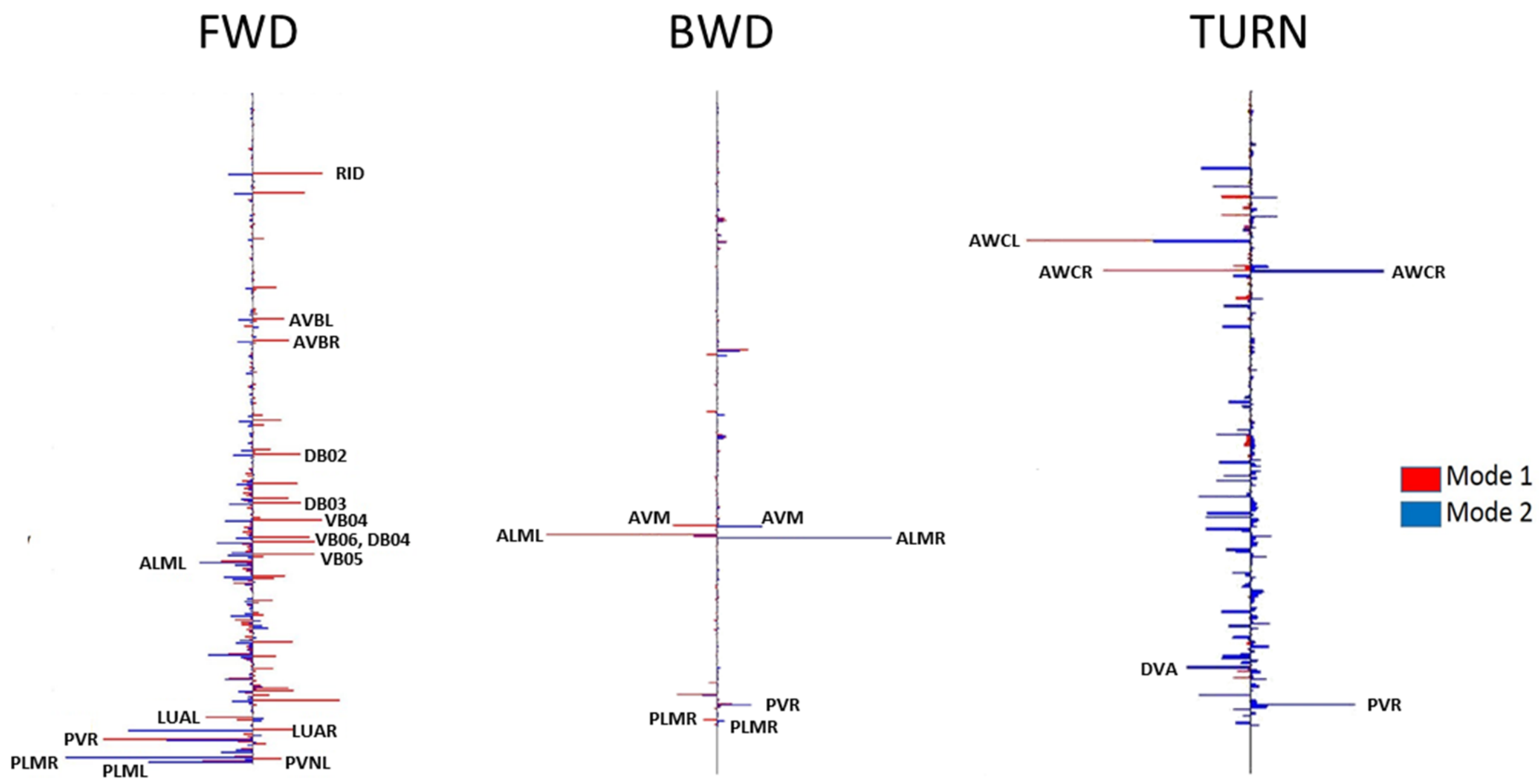

**Figure S3:** Pattern values for the first two SVD modes for neurons along the body ordered from Anterior (top) to Posterior (bottom) during forward, backward, and turning movements. Neurons in the posterior region dominate during forward motion whereas backward motion appears to be dictated by a few key neurons and turning engages anterior neurons more heavily.

The SVD decomposes $D$ into $D = \sum_{l=1}^{N} u_l \sigma_l v_l^T$, where $u_l$ are the eigenvectors of $DD^T$, representing the pattern of each mode (PC modes), $v_l$ are the eigenvectors of $D^TD$, representing the time-dependent coefficients of each mode, and $\sigma_l$ are the eigenvalues of both $DD^T$ and $D^TD$, which are the singular values that act as stretching factors. We also consider the k-mode truncated decomposition of $D$ as $D_k = \sum_{i=1}^{k} u_i \sigma_i v_i^T$ in which we take the dominant $k$ modes. For stimulation that corresponds to FWD and BWD we find that only few principal component (PC) modes are significant for the decomposition: the first PC mode explains approximately 65.24% of the energy and the first two PC modes explain about 90.94% of the energy, where the energy explained by mode k is defined as $\sum_{i=1}^{k} \sigma_i^2 / \sum \sigma^2$. Since the system is dominated by these first three PC modes, we focus on a 3-mode truncated decomposition and investigate the modes and time-dependent coefficients in this decomposition. In Fig. 3, we show the projection onto the first three modes, which corresponds to time-dependent coefficients. We also plot the absolute values assigned to each neuron in the first two modes ordered along the body of the worm in Fig, S3. We then pick the five top neurons from sensory, inter and motor neurons and plot their membranes voltages in Fig. 3 and label them in Fig. S3.

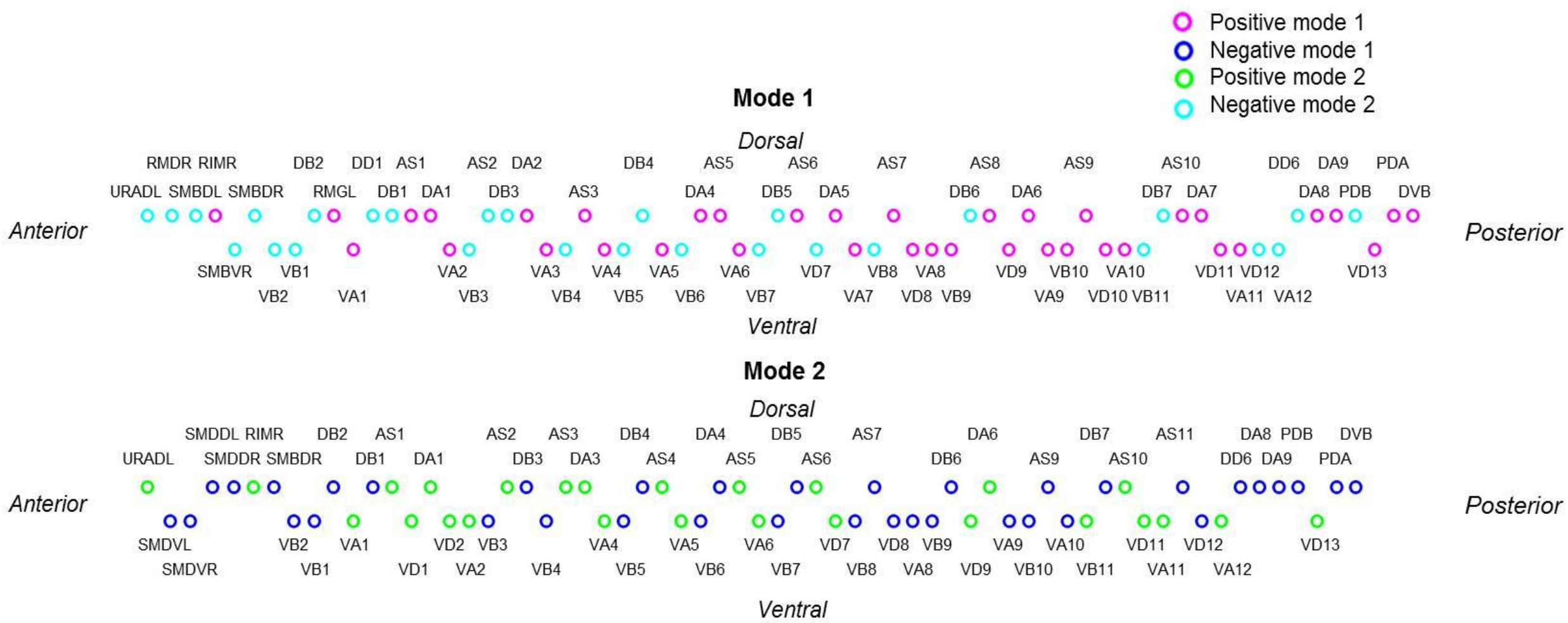

**Figure S4**: Visualization of the location and pattern values of active *C. elegans* motor neurons in the first two modes. Neurons are positioned according to their location on the AP-axis and the muscle group(s) they stimulate (dorsal or ventral). Colors indicate whether the neuron has a positive or negative pattern value in that particular mode.

In reviewing the first two modes that correspond to FWD movement in Fig. S3, it is evident that the neurons with stronger patterns values fall toward the posterior region of the body, indicating that the posterior region contains neurons that have a more significant role in *C. elegans'* forward movement. The Neural pattern plot is Fig. 3 also reveals that the ventral motor neurons (V*) have noticeably stronger pattern values than the dorsal motor neurons (D*), which implies that these neurons are responsible for a greater activity that creates the network oscillations.

In Fig. S4, we look at the pattern values for motor neurons to better identify the neurons responsible for the network behavior during forward motion. We consider a neuron "positive" or motivating forward motion if its pattern value is above the median and "negative" and less motivating if the pattern values are below the median. Using this method, we identify 32 active ventral motor neurons in each of the first two modes, 21 active dorsal motor neurons in the first mode, and 18 active dorsal motor neurons in the second mode.

In the first mode, we see 29 motor neurons that are positively activated (undulating upward), and 38 motor neurons that are negatively activated (undulating downward). In the second mode we find an opposite trend, with 37 motor neurons that are positively activated and 26 neurons that are negatively activated. Analysis of active neurons in each mode shows that there are a total of 74 active neurons, with 17 of them active only in one mode and 57 active in both modes. Notably, while such analysis can classify neurons into functional groups, it cannot directly represent the forces that will act on different segments of the body.

We also examine the first two time-dependent coefficients and see clear periodic oscillations in both. Summing up these two coefficients, we see the oscillations have a period of about 1.8 seconds and have roughly circular trajectory in the low-dimensional phase space. Such oscillations we observe in the time dependent coefficients support the correlation between the activity of the neural network and the body dynamics of the nematode; when *C. elegans* is swimming forward with its body following a periodic wave pattern, the activity of the neural network also tends to undergo periodic oscillations.

## *VI.* VISCOELASTIC ROD MODEL

In order to represent *C. elegans* movement, we apply a discrete viscoelastic rod model that has been used to describe motion in anguilliform swimmers (18). The rod representation is chosen based on the anguilliform motion seen in *C. elegans* as it reacts to various mechanical touches and similar stimuli. In this model, the swimmer's body is represented as a two-dimensional rod composed of discrete rigid segments and joints, and forces acting on the rod represent the swimmer's muscle activity. The environment of the organism is modeled through damping, which may be varied to imitate different surrounding media (18,19). The model has similar principles to the spring based biomechanical model proposed for *C. elegans* (20). We chose to work with the former since it was derived through a consistent reduction of equations describing body in fluid into discrete segments, which shown to be numerically stable and efficient under dynamic neural stimulations. The model also includes an approximation of the environment that can describe various fluid media (e.g., agar, water) with configurable properties such as fluid viscosity. Notably, our results of body movement are generic, and we do not expect them to be essentially altered with various body models.

Here we summarize shortly the modeling approach and explain how we utilize viscoelastic rod model for *C. elegans* body. The state of the viscoelastic rod at any point in time is described by the x-y coordinates of the midpoint of each segment of the rod, and by the angle, $\varphi$, of each rod segment relative to the horizontal plane. We consider the joints connecting the segments of the rod to be actuated by passive springs, dashpots, and time-dependent force generators (as visualized in Fig. 1A). Given this model, force applied near one end of the rod can lead to movement that travels through the other end of the rod, depending on the magnitude of initial force and parameters governing the rod and its environment. The equations used to calculate the position of the segments of the viscoelastic rod are:

$$x_{i+1} = \frac{h}{2}\left(cos\ (\varphi_i) + cos\ (\varphi_{i+1})\right)\ + x_i \tag{6}$$

$$y_{i+1} = \frac{h}{2}\left(sin\ (\varphi_i) + sin\ (\varphi_{i+1})\right)\ + y_i \tag{7}$$

The differential equation determining the change in $\varphi$, based on the forces **f** and **g** applied to the segments of the rod in the x and y directions, respectively, is:

$$J\ddot{\varphi} = M_i - M_{i-1} + \frac{h}{2}(g_i + g_{i-1})\cos(\varphi) - \frac{h}{2}(f_i + f_{i-1})sin(\varphi) \tag{8}$$

The components of Eqn. (8) are defined as the contact moment $M_i$ (Eqn. (9)), the moment of inertia $J_i$ for link *i* (Eqn. (10)), and the moment of inertia *I* for motions in the x-y plane (Eqn. (11)), as:

$$M_i = EI_i\left(\frac{(\varphi_{i+1} - \varphi_i)}{h} - k_i\right) + \delta_i\left(\frac{(\dot{\varphi}_{i+1} - \dot{\varphi}_i)}{h}\right) \tag{9}$$

$$J_i = \ \rho h\left(I_i + \frac{\pi}{12}r^2h^2\right) \tag{10}$$

$$I = \frac{\pi D^4}{64} \tag{11}$$

In Eqn. (9), we solve for the contact moment based on the preferred curvature of the rod $k_i$, the rod's elasticity *E* (Young's modulus), the environmental damping coefficient $\delta$, and the segment lengths $h_i$. The parameters of equations defining the moments of inertia *J* and *I* Eqn. (10, 11) depend on the rod's material density $\rho$, rod radius *r*, segment length $h_i$ and rod diameter $D$.

Expanding Eqn. (8) by substituting in Eqn. (9 - 11) we solve for $\ddot{\varphi}_i$ as follows:

$$J_i\ddot{\varphi} = M_i - M_{i-1} + \frac{h}{2}(g_i + g_{i-1})\cos(\varphi_i) - \frac{h}{2}(f_i + f_{i-1})sin(\varphi_i) \tag{12}$$

$$\rho h\left(I_i + \frac{\pi}{12}r^2h^2\right)\ddot{\varphi} = \ EI_i(\frac{(\varphi_{i+1} - \varphi_i)}{h} - k_i) + \delta_i(\frac{(\dot{\varphi}_{i+1} - \dot{\varphi}_i)}{h}) - EI_{i-1}(\frac{(\varphi_i - \varphi_{i-1})}{h} - k_i)$$
$$-\delta_{i-1}\left(\frac{(\dot{\varphi}_i - \dot{\varphi}_{i-1})}{h}\right) + \frac{h}{2}(g_i + g_{i-1})cos(\varphi_i) - \frac{h}{2}(f_i + f_{i-1})sin(\varphi_i) \tag{13}$$

$$\ddot{\varphi}_i = EI_i(\frac{(\varphi_{i+1}-\varphi_i)}{h} - k_i) + \delta_i(\frac{(\dot{\varphi}_{i+1}-\dot{\varphi}_i)}{h}) - EI_{i-1}(\frac{(\varphi_i-\varphi_{i-1})}{h} - k_i) - \delta_{i-1}(\frac{(\dot{\varphi}_i-\dot{\varphi}_{i-1})}{h})$$
$$+\frac{h}{2}(g_i+g_{i-1})cos(\varphi_i) - \frac{h}{2}(f_i+f_{i-1})sin(\varphi_i)\cdot\frac{1}{\rho h\left(I_i+\frac{\pi}{12}r^2h^2\right)} \quad (14)$$

We then transform Eqn. (14) into a system of first order differential equations:

$$p_1 = \varphi;\ p_2 = \dot{\varphi};\ \dot{p}_1 = \dot{\varphi} = p_2 \quad (15)$$

$$\dot{p}_{2i} = \ddot{\varphi}_i = \{EI_i\left(\frac{\left(p_{1(i+1)}-p_{1i}\right)}{h} - k_i\right) + \delta_i\left(\frac{\left(p_{2(i+1)}-p_{2i}\right)}{h}\right)$$
$$-EI_{i-1}\left(\frac{\left(p_{1i}-p_{1(i-1)}\right)}{h} - k_i\right) - \delta_{i-1}\left(\frac{\left(p_{2i}-p_{2(i-1)}\right)}{h}\right) + \frac{h}{2}(g_i+g_{i-1})\cos(p_{1i})$$
$$-\frac{h}{2}(f_i+f_{i-1})\sin(p_{1i})\cdot\frac{1}{\rho h\left(I_i+\frac{\pi}{12}r^2h^2\right)} \quad (16)$$

which can be solved computationally. We then include calcium dynamics defined in (18),

$$\ddot{\beta} + c_1\dot{\beta} + c_2\beta = c_3u(t) \quad (17)$$
$$\ddot{\eta} + \dot{\eta}c_4 + c_5\eta = c_6\beta(t) \quad (18)$$
$$A(t) = \frac{a_0 + (\rho\eta)^2}{1+(\rho\eta)^2} \quad (19)$$

where $u(t)$ is each motor neuron output in terms of $\beta$, the T-tubuli depolarization response and $\eta$, the SR calcium release, with constant parameters $c_1 = 60$, $c_2 = 20$, $c_3 = 50$, $c_4 = 10$, $c_5 = 30$, $c_6 = 30$. SR calcium release $\eta$ is used to define $A(t)$, the activation function with $\rho(t) = 1$, representing muscle fatigue.

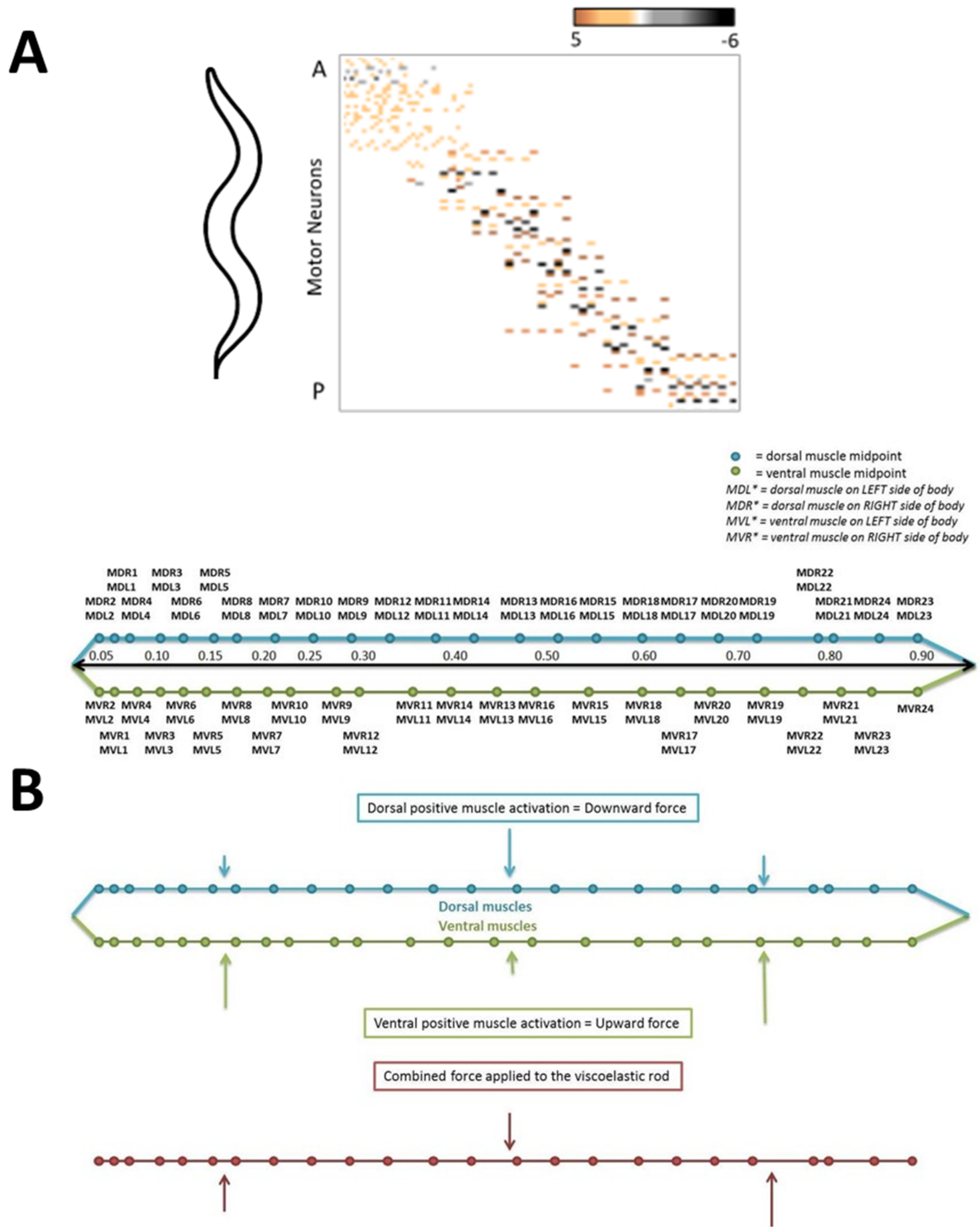

**Figure S5: A:** Top: Connectivity from motor neurons to muscles ordered from anterior (top) to posterior (bottom) direction. Bottom: Map of the 95 muscles represented as segments of the viscoelastic rod. Segment length is determined by position of each muscle relative to the AP-axis, with the overall body length scaled to 1. **B**: Force is applied to segments of the viscoelastic rod to represent muscle activation. Dorsal and ventral muscles are combined based on their locations relative to the AP-axis and summing the input into each group of muscles determines the overall force applied to a segment of the rod.

## *VII.* VISCOELASTIC ROD AS A MODEL FOR *C. ELEGANS* BODY

We apply the viscoelastic rod model to *C. elegans* by representing each muscle group in the nematode as a segment of the rod connected with joints. We approximate the forces applied to the rod segments using neuron activity data from forward motion simulations and a neuron-to-muscle map which are experimentally determined (21). The combined activity of the neurons connected to a muscle is considered the muscle's input, which leads to activation of force.

To represent accurate *C. elegans* musculature, we model 95 muscles relevant to *C. elegans* locomotion and use their approximate sizes and locations to determine segment length and position. These 95 individual muscles are divided into four groups based on their physical location in the nematode: dorsal left (DL), dorsal right (DR), ventral left (VL), and ventral right (VR). Every muscle is assigned a position along the Anterior-Posterior axis (AP-axis) of the nematode, and the dorsal left and dorsal right muscles are coupled into single segments representing the dorsal muscle groups, and likewise for the ventral muscles. Dorsal (ventral) left muscles are paired only with dorsal (ventral) right muscles where left-right pairs are determined from muscle location along the AP-axis, as shown in Fig. S5A.

Differentiation between the dorsal and ventral muscle groups is necessary since these two muscle groups act in opposition to each other; for forward motion to occur, the dorsal muscles contract while the ventral muscles relax, and vice versa. As such, we model activation of the muscles in the dorsal group through downward force, and the muscles in the ventral group through upward force. We then merge the dorsal and ventral muscle groups based on their location along the AP-axis to form a single discrete rod. With such representation, the dorsal (downward) and ventral (upward) forces can be summed over the dorsal-ventral muscle pairs to determine the overall force applied to each segment of the rod: for example, if a large downward force is applied by a dorsal muscle and a small upward force is applied by the corresponding ventral muscle, the net result will be application of downward force on the segment of the rod representing the merged muscles (Fig. S5B).

Unlike the generic anguilliform swimmer model, the basic *C. elegans* viscoelastic rod representation assumes force is only applied in the y-direction, meaning **f** is a zero vector. Given the absence of force in the x-direction, we eliminate the sine term from Eqn. (18), resulting in the final system of equations:

$$\dot{p}_{1i} = \dot{\varphi}_i = p_{2i} \tag{20}$$

$$\dot{p}_{2i} = \ddot{\varphi}_i = EI_i\left(\frac{\left(p_{1(i+1)} - p_{1i}\right)}{h} - k_i\right) + \delta_i\left(\frac{\left(p_{2(i+1)} - p_{2i}\right)}{h}\right) - EI_{i-1}\left(\frac{\left(p_{1i} - p_{1(i-1)}\right)}{h} - k_i\right) - \delta_{i-1}\left(\frac{\left(p_{2i} - p_{2(i-1)}\right)}{h}\right) + \frac{h}{2}(g_i + g_{i-1})\cos(p_{1i}) \cdot \frac{1}{\rho h\left(I_i + \frac{\pi}{12}r^2h^2\right)} \tag{21}$$

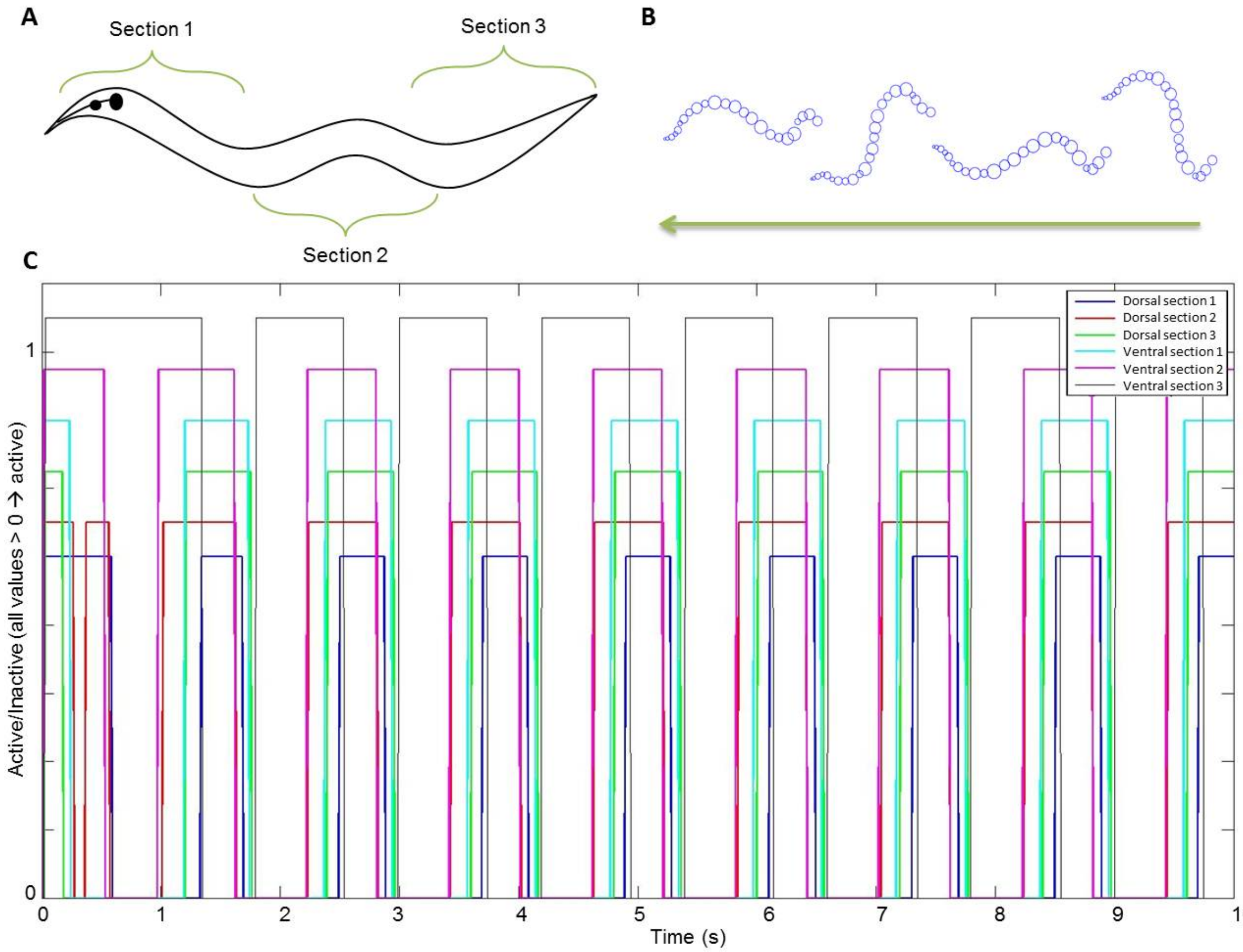

**Figure S6: A**: Three main contraction/relaxation sections. It is expected that dorsal and ventral muscles will not continuously be activated to contract simultaneously and that the stimulation of the three sections of the worm will be staggered **B**: Results of the simulations of viscoelastic rod movement over time during FWD stimulation. The worm is moving from left to right, with the size of each rod segment represented by the diameter of each circle. **C**: Activation patterns of the groups of dorsal/ventral muscles based on input from connected motor neurons during FWD excitation showing the repeating activation pattern of the six muscle groups.

for modeling *C. elegans* as a viscoelastic rod.

The parameters defining the remaining physical characteristics of the rod were set based on published analyses of high-speed images of *C. elegans* forward-swimming in a controlled environment (22) and micropipette deflection experimentation on anesthetized *C. elegans* (23). We model an adult *C. elegans* with diameter D=65μm, Young's modulus E=3.77-8 kPA and material density ρ=1.0 $g/(cm^3)$. We assume the preferred curvature to be the equilibrium state of the nematode, represented by the zero vector, and assign damping coefficient δ = 1 Ns/m as an approximate representation of the substrates making *up C. elegans* habitat (broadly ranging from rotting vegetation to animal intestines). As previously described, we assign segment lengths to vector $h_i$ based on approximate muscle length. These body-environmental parameters are fully configurable through modWorm.

The musculature and body model dynamics are computed from the normalized membrane voltages of motor neurons (i.e. $\overline{V_i} = V_i - V_{th}$). Since the system described by viscoelastic rod model is non-stiff, it can be solved using explicit integration method. We use RK4 explicit ODE solver for Python and DP5 solver for Julia. **Computational time is optimized to match real-time with a ratio of 1:1**, i.e., 1 sec of integrated body dynamics ~ 1 sec of actual body dynamics. Notably, the simulation time step for body is set to match that of nervous system dynamics, thus allowing simulation of both neural and body dynamics in the same time sampling frequency. Like neural simulations, multiple body simulations can be run in parallel utilizing modWorm's ensemble simulation function.

The behavior of the viscoelastic body model is validated by the known contraction and relaxation patterns found in the nematode's body during forward motion. These patterns closely align with the simulated formations of the viscoelastic rod as it moves over time. The behavior of the viscoelastic rod as it models step input currents for FWD, BWD and TURN scenarios shows a clear relationship between stimulation of motor neurons along the body and muscle movement (Fig. S7). Since the model is sensitive to the magnitude of motor neuron stimulation a muscle receives, we observe that greater force is applied in certain sections of the worm over time, creating the undulating form associated with *C. elegans* forward motion. Specifically, we often see the anterior and posterior sections receiving force in the opposite direction of the force applied to the middle section of the worm (Fig. S6B). For example, FWD motion configuration represents dorsal (ventral) contraction in the anterior and posterior sections with ventral (dorsal) contraction in the middle section, which is consistent to the expected muscle usage.

### *C. ELEGANS* BODY SHAPES DURING MOTION

| Time | FWD | BWD | TURN |
|---|---|---|---|
| 0.05 | 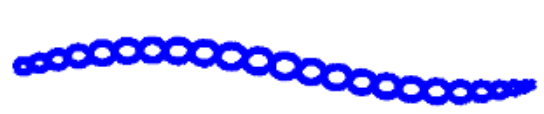 | 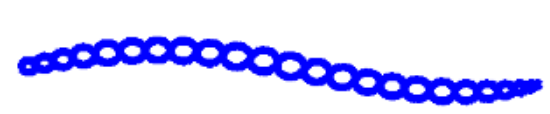 | 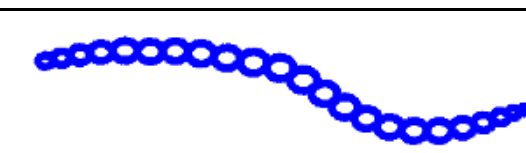 |
| 0.15 | 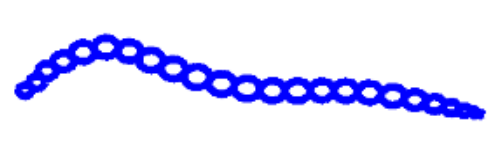 | 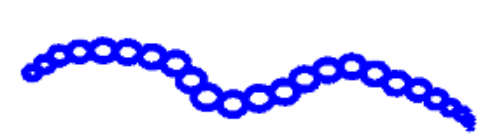 | 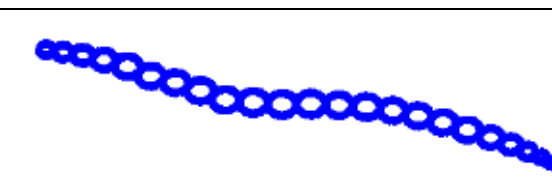 |
| 0.25 | 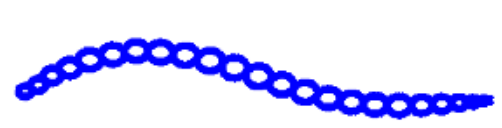 | 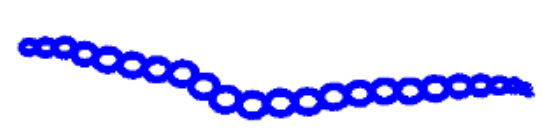 | 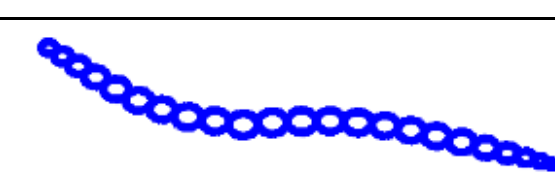 |

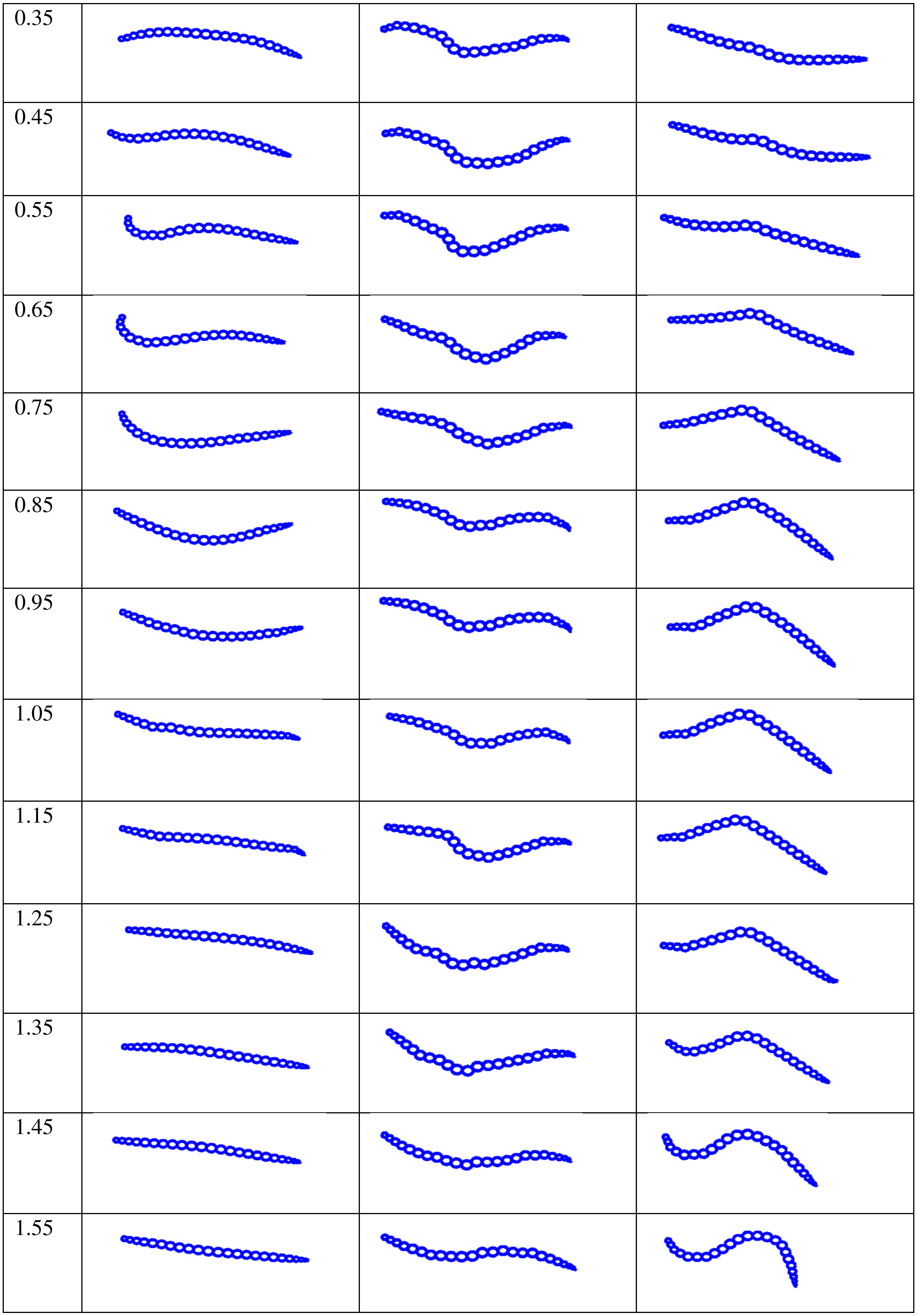
0.35
0.45
0.55
0.65
0.75
0.85
0.95
1.05
1.15
1.25
1.35
1.45
1.55

| | | | |
|---|---|---|---|
| 1.65 | 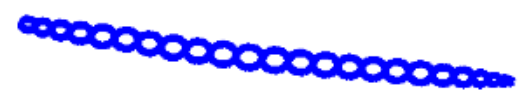 | 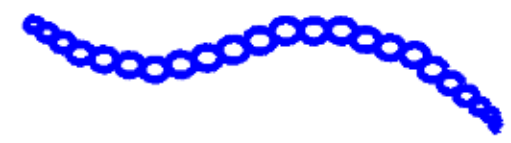 | 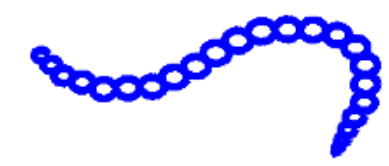 |
| 1.75 | 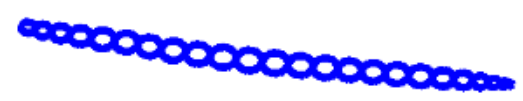 | 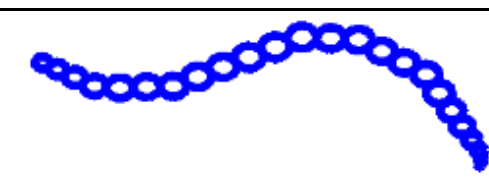 | 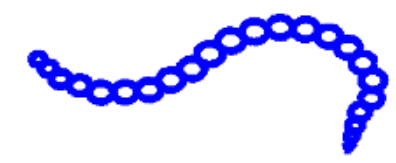 |
| 1.85 | 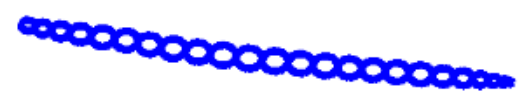 | 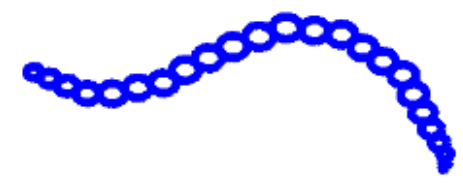 | 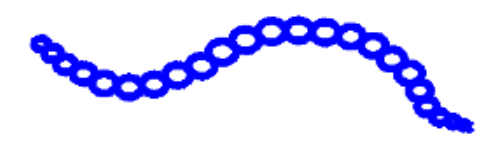 |
| 1.95 | 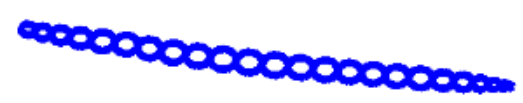 | 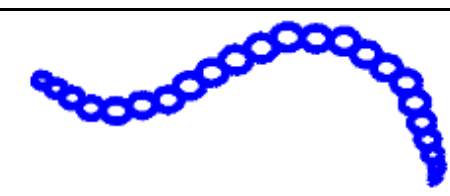 | 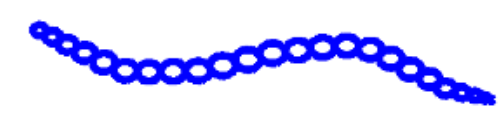 |
| 2.05 | 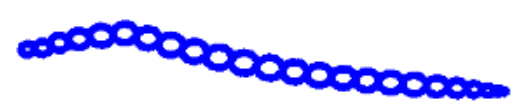 | 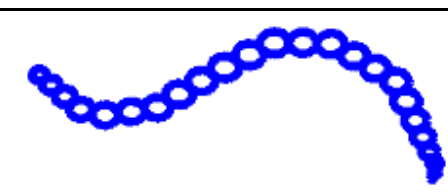 | 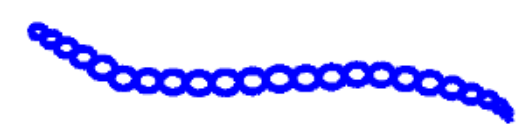 |

**Figure S7:** Library of body shapes seen during simulation of forward, backward, and turning motion induced by constant neural stimuli. Backward and turning shapes are more extreme than the smooth shapes associated with forward motion.

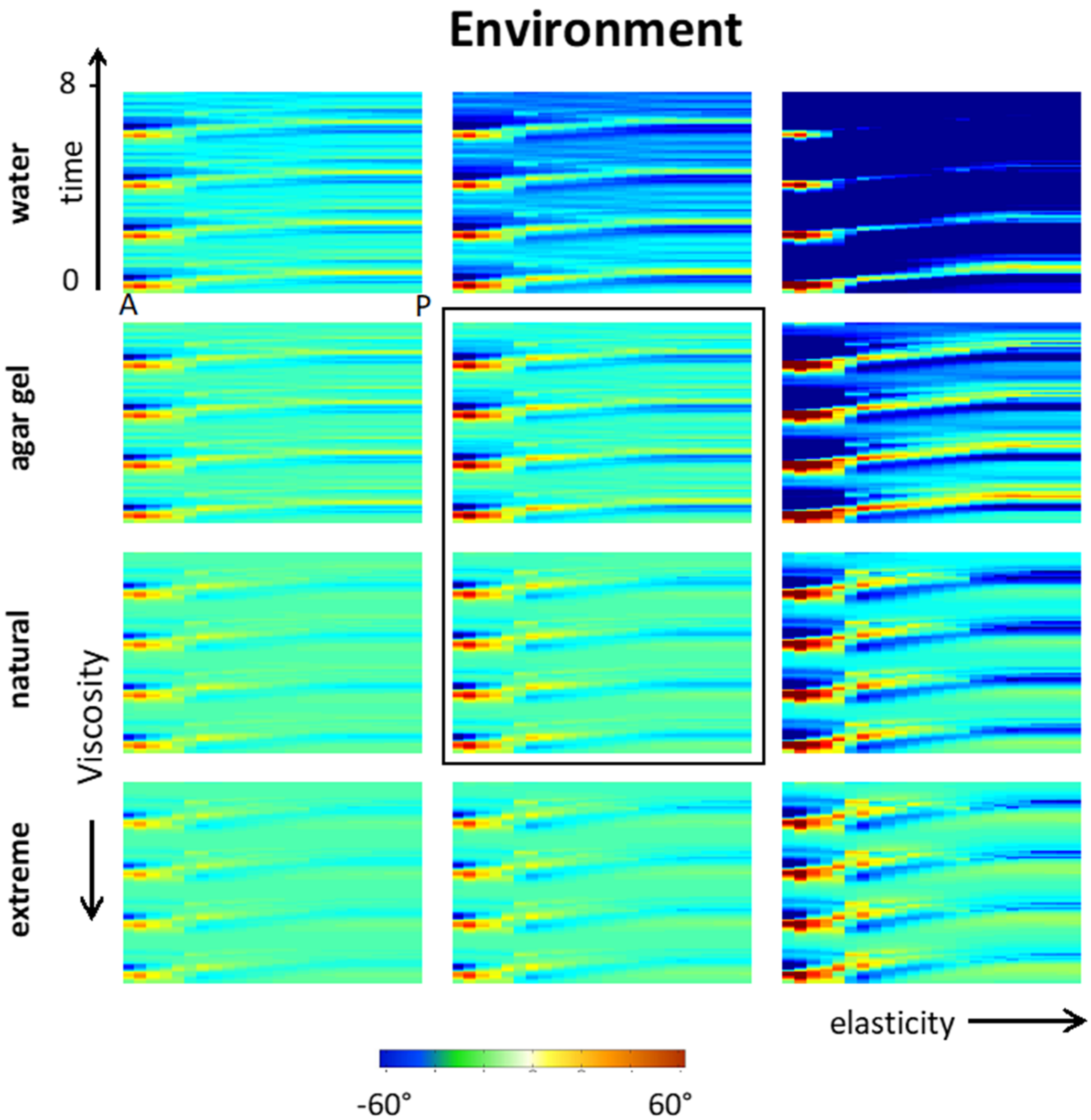

**Figure S8:** Effect of viscosity variation and body elasticity on forward movement invoked by constant stimulus (viscosity: 1, $10^2, 10^3, 10^4$ $cP$; elasticity: 3.75, 6,8. 25 kPA). Only specific choice of parameters permits coherent movement.

## *VIII.* ENVIRONMENTAL VARIATIONS

Experiments indicate that the environment plays a significant role in shaping coordinated movement (24,25). In Fig. S8, we vary fluid viscosity and body elasticity to study their effects on simulated body postures (i.e. curvature plots) during forward locomotion as characterized in Fig. 4, S7. Increasing viscosity values impedes movement by breaking off wave propagation from anterior to posterior, whereas decreasing them causes rapid extreme strokes unlike the body shapes seen in efficient *C. elegans* forward motion. When elasticity variations are added, these strokes intensify and create atypical movements for high elasticity or no movement for low elasticity.

## *IX.* BACKWARD INTEGRATION FROM MUSCLE FORCE TO NEURAL VOLTAGE

We add backward integration to our model to approximate muscles to neural dynamics interaction in addition to neurons stimulating muscles (26,27). The procedure is as follows: for an arbitrary external force acting on muscles at time *t*, the force can be approximated in the form of activation signals to muscles described as column vectors $\vec{E}_D(t), \vec{E}_V(t)$ each with dimension $(24 * 1)$ for dorsal and ventral segments of the body respectively. These vectors are then expanded into $\vec{E}_D^{Left}(t), \vec{E}_D^{Right}(t), \vec{E}_V^{Left}(t), \vec{E}_V^{Right}(t)$ where:

$$\vec{E}_D^{Left}(t) = \vec{E}_D^{Right}(t) = \vec{E}_D(t) * \frac{1}{2} \tag{22}$$

$$\vec{E}_V^{Left}(t) = \vec{E}_V^{Right}(t) = \vec{E}_V(t) * \frac{1}{2} \tag{23}$$

These 4 vectors account for the full profile of forces throughout the dorsal, ventral, left and right axis of the body where left and right forces are assumed to be symmetric in the direction of left and right axis of the body. $\vec{E}_D^{Left}(t), \vec{E}_D^{Right}(t), \vec{E}_V^{Left}(t), \vec{E}_V^{Right}(t)$ are then vertically stacked to a single column vector $\vec{E}_{DV}(t)$ of dimension $(96 * 1)$, which describes the muscle forces for the whole body of the worm at time t.

We then solve the inverse problem of forward integration process (i.e. neural dynamics to muscle forces). During the forward integration, muscle forces at time $t$ is the linear mapping of neural voltages $V(t)$ (279 * 1) via muscle mapping matrix $M(96 * 279)$:

$$\vec{E}_{D,V}(t) = MV(t) \tag{24}$$

Matrix $M$is of *fat* type (i.e. more columns than rows), meaning that for a given force $\vec{E}_{D,V}(t)$, there exist multiple neural voltage solutions which satisfy Eqn. 24. But since $M$ is not of full rank, we cannot solve $V(t)$ with ordinary least squares method. This can be circumvented using generalized pseudo-inverse conceived by Singular Value Decomposition. Using this technique, we approximate the neural voltages from muscle forces by solving the following inverse equation.

$$\vec{V}_E(t) = M^{+}\vec{E}_{D,V}(\mathrm{t}) \tag{25}$$

Here, $M^{+} = \mathrm{V}\mathrm{s}^{-1}\mathrm{U}^{\mathrm{T}}$ is the generalized pseudo-inverse of $M$ computed via taking the inverse of Singular Value Decomposition of $M$ where U and V correspond to unitary matrices each containing left and right singular vectors of M and s is the diagonal matrix whose diagonal entries are the singular values of M. Notably, Eqn. 25 effectively solves for the least norm solution $\vec{V}_E(t)$ that satisfies

$$\left|M\vec{V}(t) - \vec{E}_{D,V}(t)\right|_2 \geq \left|M\,\vec{V}_E(t) - \vec{E}_{D,V}(t)\right|_2 \tag{26}$$

i.e., $\vec{V}_E(t)$ is the least norm solution which minimizes the Euclidean norm of error (28).

The approximated voltage $\vec{V}_E(t)$ can be interpreted as the inferred neural voltage at time *t* induced by muscle forces $\vec{E}_{D,V}(t)$. To drive the nervous system with $\vec{V}_E(t)$, we define the superposed voltage:

$$V_i^{sp} = V_i + V_{E_i} \tag{27}$$

where $V_i^{sp}$ is the voltage state of ith neuron superposed with $V_{f_i}$. Equivalently, Eqn. 27 can be written as a current term induced by muscle force $I_i^{force}(\boldsymbol{\vec{V}}_{E_i})$. The modified equation allows driving the nervous system entirely by muscle to neural dynamics interactions. Specifically, the network dynamics driven by $\vec{V}_E$ is given by:

$$C\dot{V_i}(t) = -G^C\left(V_i^{sp} - E_{cell}\right) - I_i^{Gap}\left(\boldsymbol{\vec{V}}^{sp}\right) - I_i^{Syn}\left(\boldsymbol{\vec{V}}^{sp}\right) \tag{28}$$

where

$$I_i^{Gap} = \sum_j G_{ij}^{g}\left(V_i^{sp} - V_j^{sp}\right) \tag{29}$$

and

$$I_i^{Syn} = \sum_j G_{ij}^{s} s_j\left(V_i^{sp} - E_j\right) \tag{30}$$

$$s_i = a_r\phi(V_i;\beta,V_{th})(1 - s_i) - a_d s_i \tag{31}$$

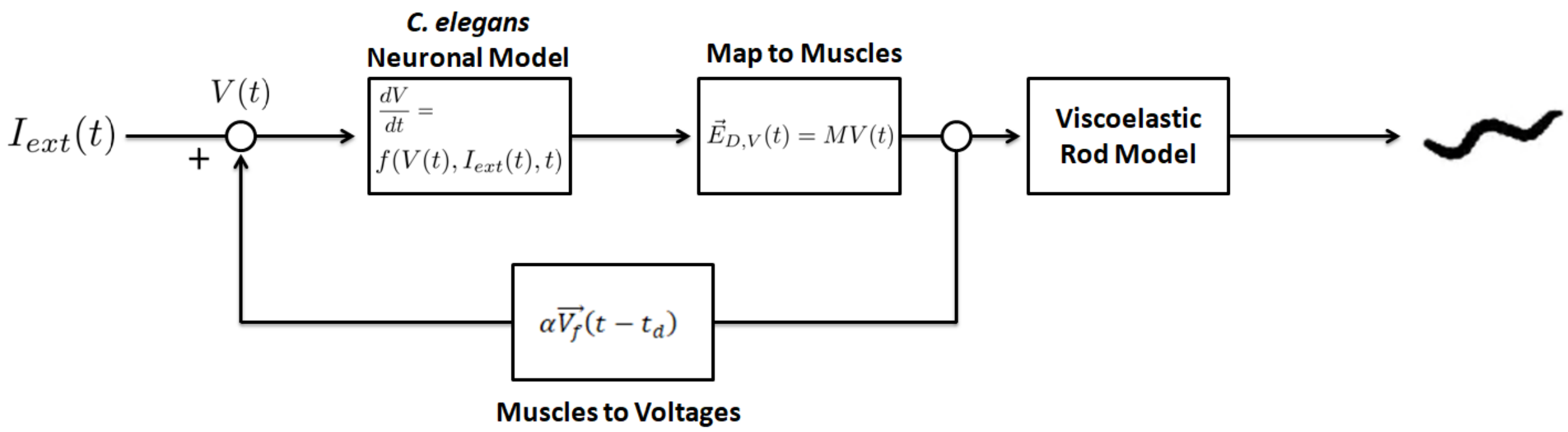

**Figure S9:** Block diagram of proprioceptive feedback implementation. $I_{ext}(t)$ is the input stimuli into nervous system at time *t*, $V(t)$ is the neuron voltages at time *t*, $\vec{E}_{D,V}$(t) is muscle forces at time *t*, and M is the muscle mapping matrix.

$$\phi(v_i; \beta, V_{th}) = \frac{1}{1 + \exp\left(-\beta\left(V_i^{sp} - V_{th}\right)\right)} \tag{32}$$

The solution $V(t)$ then describes the neural activity driven by $\vec{V_E}$ which satisfy the right-hand side of Eqn. 28. Notably, Eqn. (28 - 32) are nonlinear and therefore computed neural voltage at $t + \Delta t$ transformed to muscle force is not guaranteed to be identical to the initial arbitrary muscle force ($\vec{E}_D(t), \vec{E}_V(t)$). Such selectivity of force by the nervous system is best described in Fig. 3. The difference between initial force and output force by the nervous system can also occur when the external neural stimulation occurs simultaneously with external muscle force stimulations.

## *X.* PROPRIOCEPTIVE FEEDBACK

We use muscle forces to neurons inverse integration to emulate proprioceptive feedback from the environment. We assume that when the muscles exert forces on the environment, there are reactive forces from the particles in the local environment that are proportional to the muscle forces. Such interaction is particularly evident in distinct locomotion patterns of *C. elegans* in different fluids such as agar vs water (25). We model the environment-body interaction through an approximation in which environmental reactive forces are sensed by the nervous system with a particular delay. We implement the delayed feedback as follows: at each time $t$ muscle forces are transformed to voltages $\overrightarrow{V_E}(t)$ according to Eqn. (25). Once simulation has passed certain time $t > t_{init}$, e.g., duration of initial pulse of input stimuli associated with a mechanical touch, the voltages $\vec{V}(t)$ in Eqn. (1 - 5) is set to be superposition of intrinsic voltages $\vec{V}$ and voltages $\overrightarrow{V_E}$ induced by delayed proprioceptive feedback:

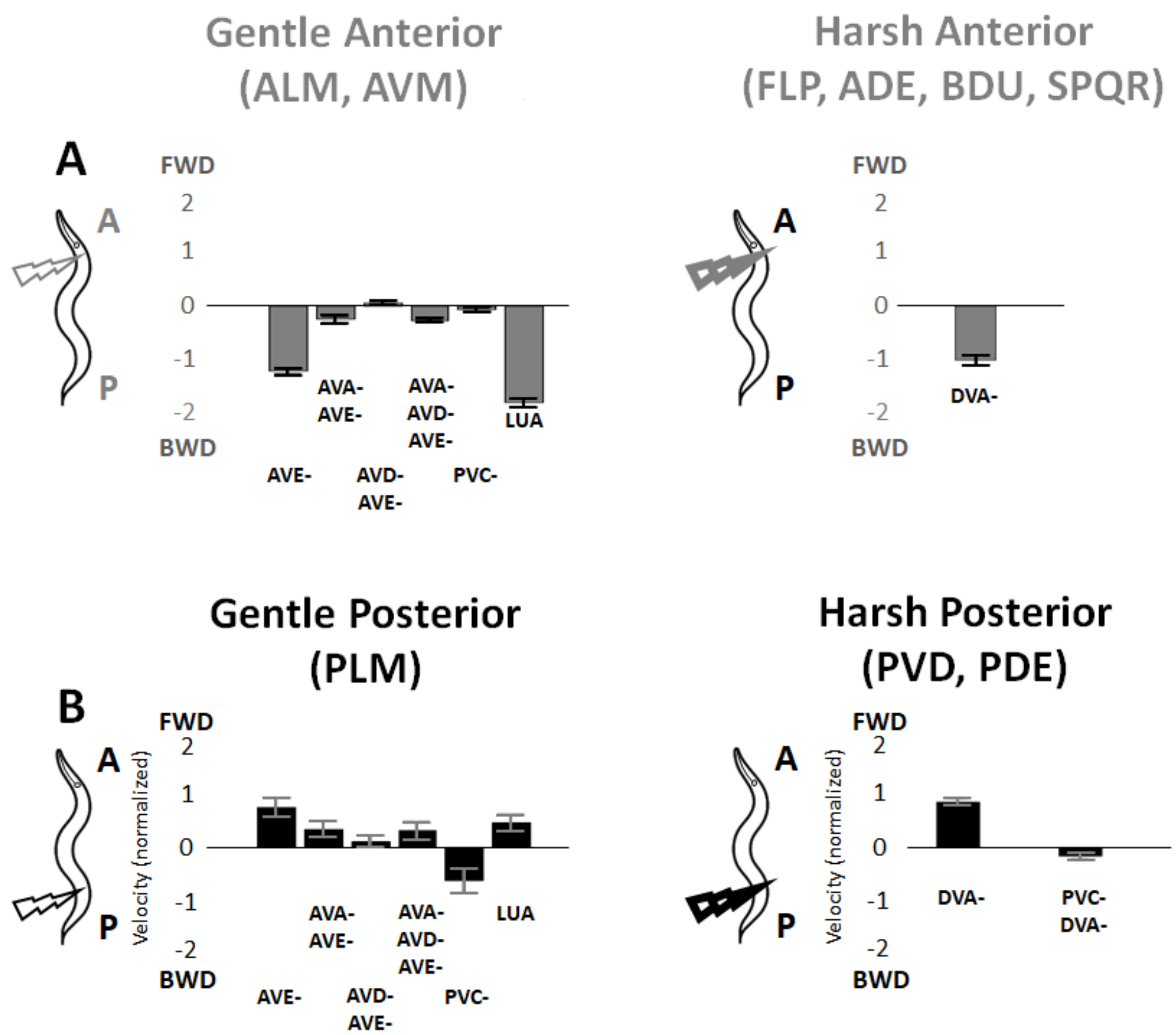

**Figure S10**: Additional *in-silico* ablations for gentle and harsh touch responses. **A:** Additional ablations for anterior gentle and harsh touch responses. **B:** Additional ablations for posterior gentle and harsh touch responses (Supplement to Figure 6).

$$\vec{V}(t) \rightarrow \vec{V}(t) + \alpha\overrightarrow{V_E}(t - t_d) \tag{33}$$

where $\alpha$ is the scaling factor and $t_d$ is the time delay. Equivalently, this is identical to adding a current term $I_i^{force}(\vec{\boldsymbol{V}}_{E_i})$ to the Eqn. 1. The block diagram of the implementation is shown in Fig. S9.

We show in Fig. 4 the effect of the delay on the curvature and find optimal values. In Fig. S11, we show motor neurons voltage profiles for movements enabled by feedback such as forward, backward locomotion and composite movements such as avoidance. Movements are triggered by short pulse of neural stimuli, which are gradually turned off exponentially, and sustained by feedback.

## *XI.* OPTOGENETIC ABLATION OF AVA

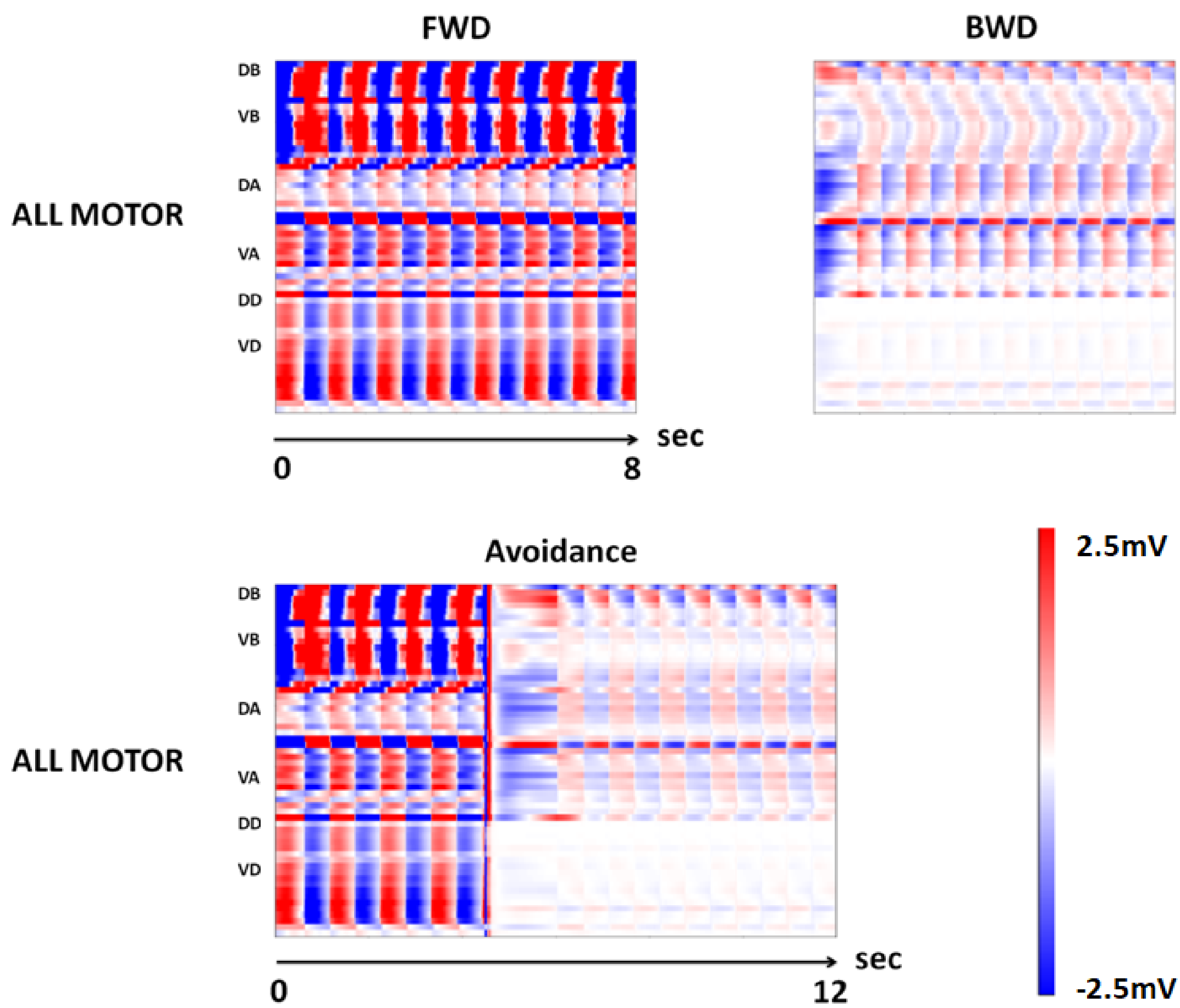

**Figure S11**: Motor neurons dynamics for forward, backward, and avoidance behaviors with proprioceptive feedback (supplement to Figure 7).

Standard culture methods were used to maintain strains (29). Animals were grown on nematode growth medium (NGM) plates and maintained on a diet of the *E. Coli* strain OP50. The N2 Bristol strain was used as the wild-type control. The strain ZM7198 (*lin-15(n765ts); hpEx3072*[*Prig-3*-MiniSOG::SL2::wCherry] +pL15EK[*lin-15AB* genomic DNA]) was used for AVA ablation (30). All strains were maintained at a low population density in dark conditions.

**AVA ablation.** We used the strain ZM7198 expressing the miniature singlet oxygen generator (miniSOG) in AVA neurons under control of the *rig-3* promoter and exposed animals to blue light to kill neurons through the production of reactive oxygen species, as previously described (31). Briefly, 20-30 larval stage 4 (L4) animals were washed in M9 buffer to remove residual food and placed on an unseeded NGM plate surrounded by a ring of 150 mM $CuSO_4$ to prevent worms from crawling away during illumination. Blue light was delivered for 12 minutes using a 488 nm 50 W LED (Mightex Systems) at a 50% duty cycle with a pulse width of 0.25 seconds using a Pulser USB light train generator (Prizmatix,

Ltd.). The plate was positioned at approximately 15 cm from the light source such that the intensity of light delivered was 200 $mW/cm^2$ as measured by a PM100D optical power meter (Thorlabs, Inc.). Animals were allowed to recover for at least 1 hour in the dark before behavioral assays. AVA ablation was verified by uncoordinated backing in response to anterior touch (32). For control experiments wild-type N2 animals were subjected to the same light stimulation protocol described above.

***In-vivo* behavioral assays.** Animals that had been allowed to recover after illumination were transferred to 6 cm NGM plates with a thin bacterial lawn. Fresh plates were made daily by spreading 30 µl of OP50 evenly across the surface of the plate and incubated at room temperature overnight. Touch assays were performed as previously described (32). Briefly a sable hair from a paint brush was taped to a glass pipette and used to gently touch animals near the tail. The video was recorded for up to a minute following touch or until the animal left the field of view. Behavioral data were aligned relative to the time of touch for comparison. Video recordings were made on a Nikon SMZ800 trinocular microscope (Nikon Instruments, Inc.) with a Pike F421b digital camera (Allied Vision Technologies, GmbH) using Fire-I 6.0 image acquisition software (Unibrain, Inc.) at a rate of 15 frames per second. Videos were analyzed using WormLab 3.1 worm tracking software (MBF bioscience) to obtain velocity, bending amplitude and position data for individual animals. Graphs were generated using Matlab R2019a (The Mathworks, Inc.) and Graphpad Prism 8.2.1 (GraphPad Software, Inc.).

## *XII.* COMPUTING EIGENWORM COEFFICIENTS FROM SIMULATIONS

Recalling from section VI, the viscoelastic model solves the segment angles $\varphi_i$ ($i \in [1,..,24]$) at each timepoint alongside the body in respect to the horizontal line in x-y plane. Since eigenworm coefficients assume there are 48 such angles along the body, we extrapolate $\varphi_i$ to span $i \in [1,..,48]$. After extrapolation, we normalize $\varphi_i$ in respect to mean angle $\overline{\varphi_i}$ for each timestep and project experimentally obtained eigenworm modes via performing dot product between $\varphi_t$ and $M$ where $M$ is (48 * 7) matrix where the columns consist of first 7 normalized eigenworm modes. Once we have projections for all timepoints, we then obtain the coefficients $s_i$ for first 7 eigenworm modes by computing the 2-norm for each projected mode (i.e. column) alongside the time axis. Finally, we obtain the normalized coefficients through $s_i / \sum s_i$. The normalized coefficients error with respect to the empirical values is then computed by calculating the mean absolute error between simulated vs experimentally obtained normalized coefficients as follows:

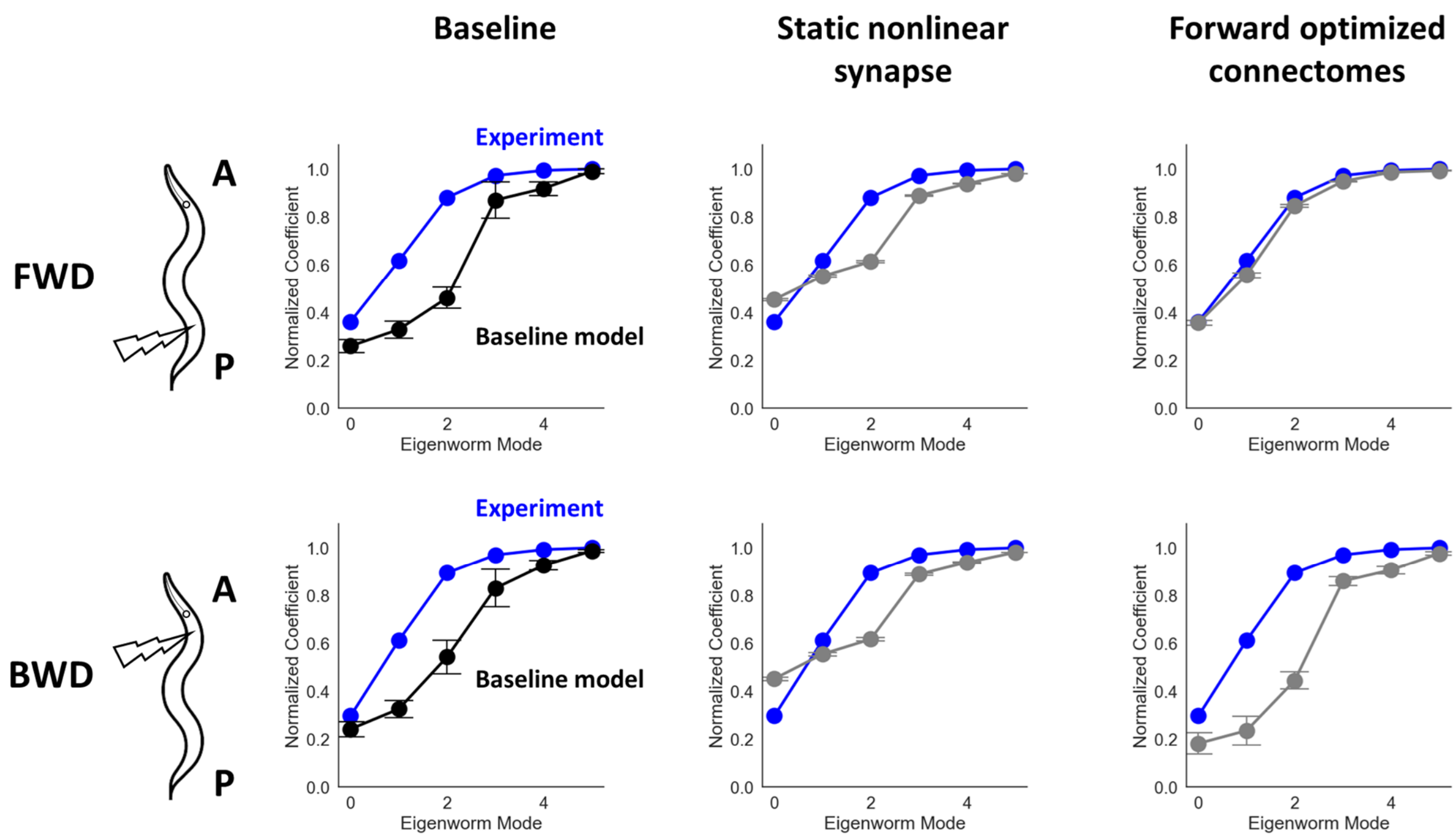

**Figure S12**: The cumulative eigenworm coefficients for forward and backward locomotion with respect to experimental coefficients (blue). From left to right, cumulative coefficients for the first 6 modes using baseline mode, baseline model using static nonlinear synapses of the form $s \sim \sigma(v)$ instead of equation 4 and using optimized connectomes with respect to forward locomotion coefficients.

$$error = \frac{1}{n}\sum_{i=1}^{n}|s_i - \hat{s}_i| \tag{34}$$

## *XIII.* CONNECTOME OPTIMIZATION

We use the genetic algorithm to optimize the synapse strengths of connectome data provided by (33) where we optimize the scaling factors $\alpha_{ij}$ to each existing synapse weight $w_{ij}$ for both gap and synaptic connectomes, amounting to total 5146 scaling factors to be optimized. The scaling factor range is set to [0.5, 1.5] and the Mean Absolute Error (MAE) formula defined by Eqn. 34 is used as an objective function. For the hyperparameters of Genetic Algorithm we use the population size of 17 with 100 generations with crossover probability of 95% and mutation probability of 70%.

## *XIV.* STATIC NONLINEAR SYNAPSES

We replace the original synaptic transmission model described by Eqn. (3, 4) with simpler static nonlinearity model using sigmoid function as follows:

$$s = \sigma(V) \tag{35}$$

$$\sigma(V) = \frac{1}{1 + e^{-\beta * V}} \tag{36}$$

Where $V$ is the neuron membrane potential and $\beta$ is the sigmoid width parameter and set to $-0.125$ identical to the value used in Eqn. 5.

## *XV.* ONLINE BLOG FOR EXPLORATIONS OF BEHAVIORAL SCENARIOS

We started a blog to share and elaborate on various stimulation and behavioral scenarios we have studied in the paper (Fig. S14). The blog will also host additional scenarios that will enhance the model and potentially contribute to better understanding of mapping neural dynamics to body movements. Through the blog we also aim to interact with the community and get it involved in identifying novel scenarios.

The blog is available as part of the Github repository:
**https://shlizee.github.io/modWorm/**

## A Initialize nervous system class

```python
class CelegansWorm_NervousSystem:

    def __init__(self):
```

Network size

```python
        self.network_Size = n_params.CE.n
```

Neurons' properties

```python
        self.neuron_C = n_dyn.init_neuron_C(capacitance = n_params.CE.cell_caps) # (n,)

        self.neuron_Linear = n_dyn.init_neuron_Linear(conductance = n_params.CE.leak_conductances,
                                                      leak_voltage = n_params.CE.leak_potentials) # (2 x n)

        self.neuron_Chemical = n_dyn.init_neuron_Chemical(synaptic_rise_time = n_params.CE.synaptic_rise_tau,
                                                          synaptic_fall_time = n_params.CE.synaptic_fall_tau,
                                                          sigmoid_width = n_params.CE.B) # (3 x n)
```

Connections + interactions properties

```python
        self.network_Electrical = n_inter.init_network_Electrical(conn_map = n_params.CE.gap_conn_2011_haspel,
                                                                  conductance_map = n_params.CE.gap_conductances,
                                                                  active_mask = np.ones(279, dtype = 'bool')) # (n x n)

        self.network_Chemical = n_inter.init_network_Chemical(conn_map = n_params.CE.syn_conn_2011_haspel,
                                                              conductance_map = n_params.CE.syn_conductances,
                                                              polarity_map = n_params.CE.ei_map,
                                                              active_mask = np.ones(279, dtype = 'bool')) # (2 x n x n)
```

## B Using new connectome mapping

New adjacency matrices for Electrical and Chemical connectomes

```python
# Initialize network and its properties
self.network_Electrical = n_inter.init_network_Electrical(conn_map = gap_conn_Cook_2019,
                                                          conductance_map = n_params.CE.gap_conductances,
                                                          active_mask = np.ones(279, dtype = 'bool')) # (n x n)

self.network_Chemical = n_inter.init_network_Chemical(conn_map = syn_conn_Cook_2019,
                                                      conductance_map = n_params.CE.syn_conductances,
                                                      polarity_map = n_params.CE.ei_map,
                                                      active_mask = np.ones(279, dtype = 'bool')) # (2 x n x n)
```

## C Incorporating non-linear neural channels (AWA)

```python
self.neuron_EGL19_AWA = n_dyn.init_neuron_Nonlinear(self, channel_type = 'EGL19_awa',
                                                    neuron_inds = [73, 82],
                                                    params_mat = EGL19_params,
                                                    added_order = 1,
                                                    initconds_mat = EGL19_initcond,
                                                    using_julia = True)

self.neuron_SHK1_AWA = n_dyn.init_neuron_Nonlinear(self, channel_type = 'SHK1_awa',
                                                   neuron_inds = [73, 82],
                                                   params_mat = SHK1_params,
                                                   added_order = 2,
                                                   initconds_mat = SHK1_initcond,
                                                   using_julia = True)
```

## D Incorporating tyramine gated chloride channels (LGC-55)

```python
self.network_Chemical = n_inter.init_network_Chemical(conn_map = n_params.CE.syn_conn_2019_nw_haspel,
                                                      conductance_map = n_params.CE.syn_conductances,
                                                      polarity_map = ei_map_LGC55,
                                                      active_mask = np.ones(279, dtype = 'bool')) # (2 x n x n)
```

New adjacency matrix describing excitatory-inhibitory mapping

**Figure S13**: Using modWorm interface to incorporate each variation to the model. **A:** Defining a nervous system class prior to the simulation. "init_" functions allow defining each nervous system module with user defined parameters. **B:** To initialize the model with non-baseline connectome data (e.g. dataset from Cook et al, 2019), simply provide NumPy arrays (279 * 279) each corresponding to desired electrical and chemical connectome mappings. **C:** To initialize the model with neurons with non-linear channels, define self.channel_name on top of baseline model to incorporate neural channels to particular set of neurons with their respective parameters. **D:** To initialize the model with modified synaptic polarities, simply replace the "ei_map" parameter of init_network_Chemical() function with new NumPy array (279 * 279) incorporating modified neuron to neuron polarities (e.g., LGC-55). (supplement to Figure 7).

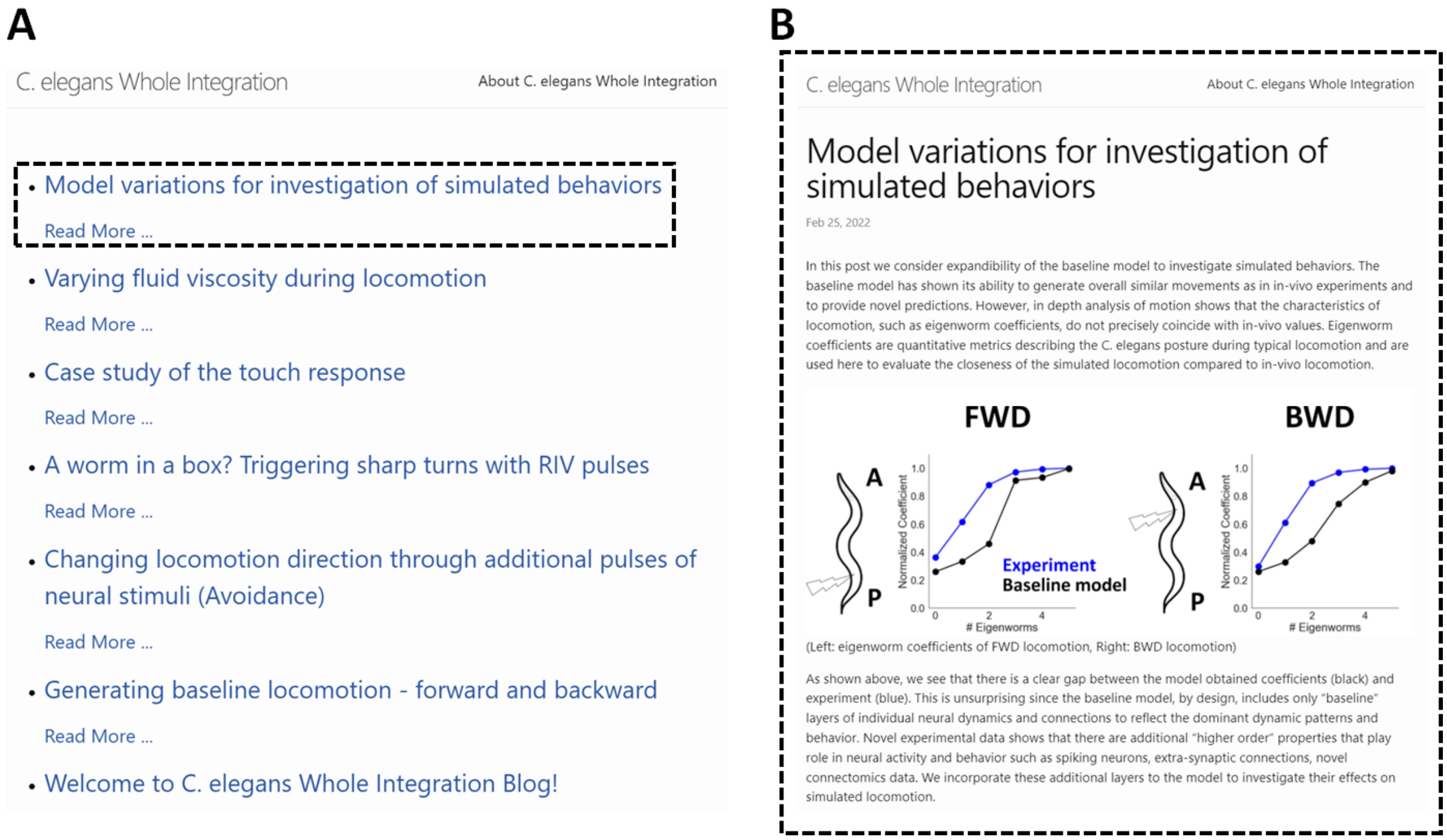

**Figure S14:** Online blog hosting the scenarios in Figure 2-7 and future additional studies. **A:** The main page of the blog featuring the studied scenarios. **B:** Redirected page after clicking on "Read More" on "Model variations for investigation of simulated behaviors" post.